\documentclass[fleqn,usenatbib]{mnras}

\usepackage[T1]{fontenc}

\DeclareRobustCommand{\VAN}[3]{#2}
\let\VANthebibliography\thebibliography
\def\thebibliography{\DeclareRobustCommand{\VAN}[3]{##3}\VANthebibliography}

\hypersetup{
    colorlinks=true,
    linkcolor=darkcolor,
    filecolor=darkcolor,
    urlcolor=darkcolor,
    citecolor=darkcolor
}
\usepackage{anyfontsize}
\usepackage{ulem}
\usepackage{caption}
\usepackage{subcaption}
\usepackage{tikz}
\usepackage{float}
\floatstyle{plaintop}
\restylefloat{table}
\usepackage[utf8]{inputenc}
\usepackage[T1]{fontenc}
\usepackage{ae,aecompl}
\usepackage{bm}
\usepackage{xcolor}
\usepackage{soul}
\usepackage{graphicx}
\usepackage{if then}
\usepackage{longtable}
\usetikzlibrary{shapes,arrows}
\usepackage{url}
\usepackage{mathtools}
\usepackage{xspace}

\usepackage{enumitem,amssymb}
\usepackage{pifont}
\definecolor{darkcolor}{rgb}{0.422,0.222,0.613} 
\definecolor{nicegreen}{HTML}{2CA02C}
\definecolor{darkgreen}{rgb}{0.0,0.5,0.0}

\newcommand{\borg}{\texttt{BORG}\xspace}

\newcommand{\fnl}{{f_\mathrm{NL}}}
\newcommand{\fnllocal}{{f^{\rm{local}}_\mathrm{NL}}}
\newcommand{\fnlgt}{{f_\mathrm{NL}^{\rm gt}}}
\newcommand{\sigmafnl}{\sigma(f_\mathrm{NL})}
\newcommand{\mvec}[1]{{\mathbf{#1}}}

\newcommand{\bpd}{b_{\phi\delta}}
\newcommand{\Mpch}{\ensuremath{h^{-1}\;\text{Mpc}}}
\newcommand{\hMpc}{\ensuremath{h\;\text{Mpc}^{-1}}}

\usepackage{newtxtext,newtxmath}

\title[]{Field-Level Inference of Primordial Non-Gaussianity with the \textit{Quijote} Simulation Suite}

\author[A. Andrews et al.]{
\parbox{2\columnwidth}{Adam Andrews$^{1,2}$\thanks{E-mail: adam.andrews@inaf.it},
Jens Jasche$^{3}$,
Guilhem Lavaux$^{4}$,
William Coulton$^{5,6}$,
Francisco Villaescusa-Navarro$^{7}$,
Marco Baldi$^{8,1,2}$,
Drew Jamieson$^{9}$,
Gabriel Jung$^{10}$,
Dionysios Karagiannis$^{11,12,13}$,
Florent Leclercq$^{4}$,
Michele Liguori$^{14,15}$,
Marco Marinucci$^{16}$,
\mbox{Benjamin}~D.~Wandelt$^{17,18,4}$\newline
}\\
\\
$^{1}$ INAF/OAS Bologna, via Piero Gobetti 101, I-40129 Bologna, Italy\\
$^{2}$ INFN, Sezione di Bologna, via Irnerio 46, I-40126 Bologna, Italy\\
$^{3}$ The Oskar Klein Centre, Department of Physics, Stockholm University, AlbaNova University Centre, SE 106 91 Stockholm, Sweden\\
$^{4}$ Institut d'Astrophysique de Paris, UMR 7095, CNRS, and Sorbonne Universit\'e, 98 bis boulevard Arago, 75014 Paris, France\\
$^{5}$ Kavli Institute for Cosmology Cambridge, Madingley Road, Cambridge CB3 0HA, UK\\
$^{6}$ DAMTP, Centre for Mathematical Sciences, University of Cambridge, Wilberforce Road, Cambridge CB3 OWA, UK\\
$^{7}$ Center for Computational Astrophysics, 160 5th Avenue, New York, NY, 10010, USA\\
$^{8}$ Dipartimento di Fisica e Astronomia, Alma Mater Studiorum - University of Bologna, Via Piero Gobetti 93/2, 40129 Bologna BO, Italy\\
$^{9}$ Max-Planck-Institut f\"ur Astrophysik, Karl-Schwarzschild-Straße 1, 85748 Garching, Germany\\
$^{10}$ Université Paris-Saclay, CNRS, Institut d’Astrophysique Spatiale, 91405 Orsay, France \\
$^{11}$ Dipartimento di Fisica e Scienze della Terra, Universit{\`a} degli Studi di Ferrara, via Giuseppe Saragat 1, 44122 Ferrara, Italy \\
$^{12}$ INFN, Sezione di Ferrara, via Giuseppe Saragat 1, 44122 Ferrara, Italy \\
$^{13}$ Department of Physics \& Astronomy, University of the Western Cape, Cape Town 7535, South Africa \\
$^{14}$ Dipartimento di Fisica e Astronomia “G. Galilei”, Universit`a degli Studi di Padova, via Marzolo 8, I-35131, Padova, Italy\\
$^{15}$ INFN, Sezione di Padova, via Marzolo 8, I-35131, Padova, Italy\\
$^{16}$ Institute for Theoretical Physics, ETH Zurich, 8093 Zurich, Switzerland\\
$^{17}$ Department of Physics and Astronomy, Johns Hopkins University, 3400 North Charles Street, Baltimore, MD, 21218, USA\\
$^{18}$ Department of Applied Mathematics and Statistics, Johns Hopkins University, 3400 North Charles Street, Baltimore, MD, 21218, USA\\
 }

\date{Accepted XXX. Received YYY; in original form ZZZ}

\pubyear{2026}

\begin{document}
\label{firstpage}
\pagerange{\pageref{firstpage}--\pageref{lastpage}}
\maketitle

\begin{abstract}
Local primordial non-Gaussianity, parameterised as $\fnllocal$, will be stringently constrained using state-of-the-art methods applied to next-generation galaxy redshift survey data. In this paper, in preparation for the upcoming data sets, we demonstrate for the first time the joint field-level inference of $\fnllocal$, nuisance parameters, and the initial conditions in realistic halo catalogues, ones which are generated through full dark-matter-only $N$-body simulations. The field-level inference algorithm optimally constrains $\fnllocal$ through a Bayesian forward-modelling approach at the field level, which outperforms traditional methods by leveraging the full statistical power of the data at the scales considered. In addition, we assess its performance under various design choices in the forward model, including tests of the structure formation model and resolution. We demonstrate the robustness of our approach by applying it to a subset of the \textit{Quijote} simulation suite, performing the inference at scales down to $k_{\rm max} \approx 0.1 \hMpc$. Compared with a power spectrum and bispectrum estimator, we find a $\sim1.3$ improvement in $\sigma(\fnllocal)$ when applying \borg{}, while marginalising over the initial conditions and bias parameters.
From the small-scale information sensitivity tests, we show that the constraints on $\fnllocal$ improve as we increase the resolution of the inference. These findings underscore the transformative potential of field-level inference to leverage the information available in ongoing surveys such as \textit{Euclid}, providing accurate insights into the physics of cosmic inflation and the number of fields driving it.
\end{abstract}


\section{Introduction}
\label{INTRO}

One of the major questions in modern cosmology is the physical mechanism that generates the initial conditions of the Universe \citep[][]{2004PhR...402..103B,biagetti_hunt_2019,achucarro_inflation_2022}. These initial conditions set the foundation for the formation of the large-scale structure we observe today. However, the actual process, or even the dominant particle(s), remains unclear. The canonical mechanism for generating these initial conditions is cosmic inflation, an era of the early universe where an accelerated expansion occurred \citep[][]{starobinsky_new_1980,guth_inflationary_1981}. In particular, vanilla models, such as the single-field slow-roll model, predict nearly Gaussian initial conditions \citep[][]{1990PhRvD..42.3936S,Gangui_1993tt,Acquaviva_2002ud,2003JHEP...05..013M,Creminelli_2004yq,byrnes_review_2010,Creminelli_2011rh,Baldauf_2011}. Consequently, detecting local Primordial Non-Gaussianity (PNG) would provide strong evidence for the involvement of multiple fields during cosmic inflation and would serve as compelling evidence for the era of cosmic inflation \citep[][]{Chen_2010,2010CQGra..27l4010K,alvarez_testing_2014,CORE_2016ymi,2018arXiv181208197C,meerburg_primordial_2019,1993ApJ...403L...1F, Gangui_1993tt,2001PhRvD..63f3002K,2003JHEP...05..013M,2004PhR...402..103B,2010AdAst2010E..72C,biagetti_hunt_2019,meerburg_primordial_2019,pardede2023wideangle,2023arXiv231104882G}.

\begin{figure}
	\centering
    \includegraphics[width=1.0\columnwidth]{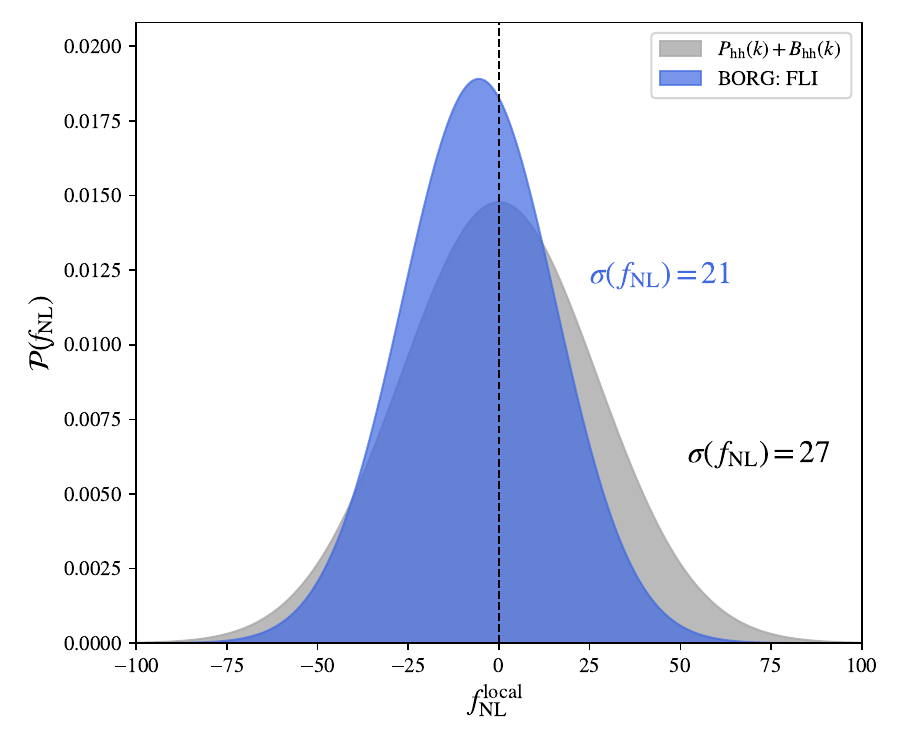}
	\caption{Comparison of constraints on $f_{\mathrm{NL}}^{\mathrm{local}}$ from two different techniques in the \textit{Quijote} simulation suite. The grey posteriors represent the Fisher constraints using a combination of power spectrum and bispectrum estimators. 
    The blue posterior shows the constraints on $f_{\mathrm{NL}}$ from a \borg{} analysis, with an uncertainty of $\sigmafnl = 23$. The mean and the uncertainty of the posterior is estimated from the weighted mean and the total standard deviation of the runs performed. The vertical dashed line indicates the fiducial value $f_{\mathrm{NL}} = 0$. The results showcase the possible constraints on $\fnl$ at the given resolution of $k_{\rm max} \approx 0.1 \hMpc$ when using a field-level inference method to the \textit{Quijote}-PNG suite over a traditional estimator. The description of the computation of the forecasted constraints is provided in Section \ref{sec:fisher_method}.}
    \label{fig:money_plot}
\end{figure}

The most stringent constraint on local PNG, parameterised as $\fnllocal$ (henceforth denoted as $\fnl$) currently comes from Cosmic Microwave Background (CMB) measurements by the \textit{Planck} satellite, reporting $\fnl = -0.1 \pm 5.0$ at a confidence level of 68.3\% \citep{2025A&A...702A.204J,planck_collaboration_planck_2019_IX}. While the CMB remains a valuable cosmological probe of the early universe, its potential for new information on PNG is limited; large-scale temperature fluctuations are expected to have reached the cosmic-variance limit. Although small-scale polarisation data may yield marginal improvements \citep{CMBPolStudyTeam_2008rgp,2020PhRvD.102b3521D,CMB_S4_2016ple}, it is unlikely to achieve the target of $\sigmafnl=1$ \citep{biagetti_hunt_2019,meerburg_primordial_2019,achucarro_inflation_2022}. In contrast, next-generation 3D galaxy surveys, which cover vast cosmic volumes and capture the largest scales of matter distribution, hold promise for providing new insights into $\fnl$ \citep{biagetti_hunt_2019,McQuinn_2021,achucarro_inflation_2022}.

The next generation of galaxy redshift surveys are underway to significantly enhance our understanding of PNG with the aim of reducing the uncertainty in $\sigmafnl$ to the order of unity \citep[][]{lsst_science_collaboration_lsst_2009,dore_cosmology_2014,amendola_cosmology_2018,2024arXiv240513491E}. In the large-scale structure, PNG leaves characteristic imprints in various cosmological observables, with notable effects including scale-dependent bias \citep[][]{2017MNRAS.468.3277B,2018MNRAS.474.2853U,mueller_optimising_2019,Karagiannis:2020dpq,Biagetti_2022,giri2023constraining,2023MNRAS.523..603R,2024arXiv240318789B,2024arXiv240317657P,2024arXiv240313985Y,2024arXiv240300490J}, small-scale signatures \citep[][]{Baldauf_2011,karagiannis_constraining_2018,friedrich_primordial_2019,Goldstein_2022,2023arXiv231012959G}, and higher-order statistical moments \citep[][]{catelan_velocity_1995,schmidt_large-scale_2010,lam_pairwise_2011,Baldauf_2011,ma_independent_2013,Tasinato_2014,biagetti_hunt_2019,baumann2021power}. These effects provide crucial information about the early universe and the processes that shaped the Universe in which we exist and observe.

Current methods for constraining $\fnl$ rely on summary statistics to extract primordial information from galaxy clustering data \citep[][]{karagiannis_constraining_2018,castorina_redshift-weighted_2019,Karagiannis_2020,mueller2021clustering,damico_limits_2022,cabass_constraints_2022,2023arXiv230915814C}. These approaches have yielded encouraging results, with studies measuring $\fnl$ down to the level of $\sigmafnl=7.4$ \citep[DESI, $p=1.6$\footnote{See \citet{2026arXiv260212357F} for an analysis of DESI quasars with other values for $p$.} for quasars,][]{2024arXiv241117623C,2025arXiv251204266C}, $-6 < \fnl < 20$ \citep[SDSS4/eBOSS,][]{2023arXiv230915814C,2025JCAP...07..043C} and  $f_{\rm NL} = -3 \pm 14$ \citep[Quaia,][]{2024ApJ...964...69S,2025A&A...698A.177B,2026arXiv260116948R,2026JCAP...02..056F} using either a power spectrum estimator, or in conjunction with a bispectrum estimator. However, they reduce the vast amount of information available in galaxy fields to a limited set of statistical moments. This compression is not sensitive to higher-order correlations or non-Gaussian features, which are relevant to constrain $\fnl$ at the required precision. As a result, they ignore a substantial amount of information, especially in the era of high-precision data expected from upcoming surveys. Overcoming these limitations requires methods that exploit the full statistical content of the data; a wide variety of statistical methods have been developed to access information beyond 2-pt statistics, e.g., higher-order statistics or\citep{karagiannis_constraining_2018,2024arXiv240318789B,2024arXiv240913583M,2024arXiv240813876W,2024arXiv241114377M,2024arXiv240300490J}, persistent homology \citep{Biagetti_2022,2024arXiv240313985Y,2025arXiv251209852C}, including machine learning approaches \citep[e.g.,][]{giri2023constraining,2023PhRvD.107f1301G,2024arXiv241001007K,2024JCAP...02..031F,2024arXiv240317657P,2024MNRAS.529.3289N,2024arXiv241200968C,2024arXiv240909124P,2025arXiv251012715C}. A prominent example is field-level inference, which preserves all available information and allows for a more direct connection between theory and observation.

\begin{figure*}
\includegraphics[width=2.0\columnwidth, trim={1.5cm 2.9cm 10cm 1.5cm},clip]{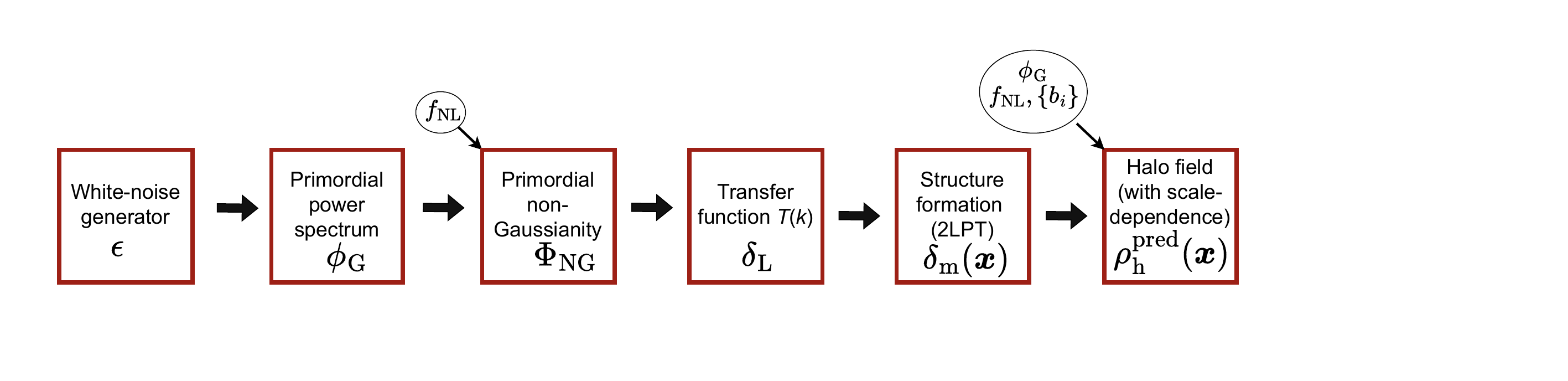}
\caption{Illustration of the forward model used in this study. The model transforms a given set of initial conditions into a predicted halo field, incorporating key features such as the perturbation of the primordial gravitational potential by $\fnl$ and scale-dependent terms in the galaxy bias model. The predicted halo field is then compared to the observed data through voxel-level likelihood evaluation. For more details, see Section \ref{Method}.}
\label{fig:flowchart}
\end{figure*}

Field-Level Inference (FLI) has found traction as an alternative to traditional methods to analyse galaxy survey data \citep[][and references therein]{jasche_bayesian_2013,jasche2017,2017JCAP...12..009S,ramanah_cosmological_2019,2023PhRvD.108h3506Z,2025MNRAS.540..716M}
Unlike conventional approaches that rely on data compression, such as power spectrum and bispectrum estimators, FLI directly fits the physics model to the observational data by forward modelling galaxy redshift data to constrain the initial conditions \citep[][]{2015JCAP...01..036J,2015arXiv151202242L,2017JCAP...06..049L,2016MNRAS.455.3169L,2019A&A...625A..64J,Lavaux2019,2025MNRAS.540..716M,2026arXiv260202363A}. This method has the potential to fully exploit the available information in galaxy observations by simultaneously reconstructing the initial conditions and constraining key model parameters, such as $\fnl$ \citep[][]{andrews_bayesian_2023,2024arXiv241018039S,andrews_epng}. FLI has been applied to, for example, weak lensing \citep[][]{2021MNRAS.502.3035P,2022MNRAS.509.3194P,2023arXiv230404785P,2024PhRvD.110b3539Z}, the constraint of photometric redshift errors \citep{2012MNRAS.425.1042J,2023arXiv230103581T}, intrinsic alignment \citep{Tsaprazi_2022}, HI intensity mapping \citep{2023PhRvD.108h3528O}, BAO \citep{2022JCAP...08..007B,2024arXiv240701524B,2025arXiv250513588B}, Lyman-$\alpha$ forest \citep{2019A&A...630A.151P,2020A&A...642A.139P}, peculiar velocity \citep{2023MNRAS.518.4191P}, performing posterior resimulations \citep{2022MNRAS.512.5823M, 2022MNRAS.509.1432S, 2024MNRAS.534.3120S, 2024MNRAS.527.1244S, 2024MNRAS.531.2213S,  2024A&A...691A.348W}, and combined with machine learning modelling \citep{2023arXiv231209271D,2024arXiv240701391D}.

Despite its promise, further testing is required, particularly in applying FLI to more realistic simulations. The underlying physics model used in this approach plays a crucial role in accurately capturing the dynamics of the early universe and the imprints left by PNG. Therefore, refining the model and validating it against increasingly realistic datasets will be essential to deploying FLI.

In this paper, we move beyond self-consistent tests -- where the forward model is used both to generate mock data and to analyse it, potentially limiting the validity and scope of the tests -- and instead apply the algorithm to $N$-body simulations \citep[][]{Coulton_2023}, which offer a more accurate representation of the complex features present in real data. This allows us to assess the algorithm's ability to constrain local PNG in a more realistic cosmological setting. It also enables a direct performance comparison with other methods; a key point of ongoing debate regarding whether field-level inference captures information beyond summary statistics \citep{2024PhRvD.109d3526C,2025arXiv250909673A,2025JCAP...09..056S, Spezzati:2025zsb}.
In earlier work, FLI has shown to be able to constrain both cosmological parameters and initial conditions in applications to halo catalogues \citep{2024arXiv240303220N,2024arXiv240701524B,2025arXiv250513588B}, and HOD galaxies \citep{2024arXiv240502252B}. In this work, our primary goal is to demonstrate that FLI can robustly constrain PNG in realistic dark matter halo simulations at the assumed resolution. Additionally, we perform a series of additional tests to evaluate the algorithm's performance across different forward models. The main result of this paper is illustrated in Figure \ref{fig:money_plot}, and an overview of the investigations is given in Table \ref{tab:overview}.

The paper is structured as follows. We present the details of the method and outline our field-level inference approach in Section \ref{Method}, beginning with an overview of the method and a description of the physics model adopted, including the bias model used and the sampling framework. The subsequent data description in Section \ref{data} outlines the \textit{Quijote} data set used in this paper. In Section \ref{results}, we present and discuss our results corresponding to each analysis setup design. We conclude the paper with a summary of our key results, a discussion of their implications, and directions for future work, which can be found in Section \ref{conclusions}.

\section{Method}
\label{Method}

This section outlines the \borg{} algorithm (Bayesian Origin Reconstruction from Galaxies) which is the methodology used to perform field-level inference to constrain $\fnl$. The \borg{} algorithm is a Bayesian hierarchical forward-modelling framework for reconstructing the initial conditions of the Universe. We begin with a general overview of the algorithm, followed by detailed descriptions of the forward model, the statistical sampling framework, and a description of the MCMC analyses.

In general, the \borg{} framework comprises two major components. The first component is the \textit{differentiable} physical forward model. The forward model is used to bring an arbitrary set of initial conditions to a corresponding set of observations, e.g. galaxy data or halo catalogue. A schematic drawing of the forward model used in this project can be found in Figure \ref{fig:flowchart}.

The full posterior of the probabilistic problem that \borg{} solves can be formulated through Bayes' theorem:
\begin{multline}
\pi\left(\epsilon, \fnl, \left\{b_i \right\}|\rho_{\rm h}^{\rm data}\right) = \pi(\fnl)\, \pi(\epsilon)\, \pi\left(\left\{b_i \right\}\right) \times \\ \times \frac{\pi\left(\rho_{\rm h}^{\rm data}|\epsilon, \fnl, \left\{b_i \right\}\right)}{\pi\left(\rho_{\rm h}^{\rm data}\right)} \, , 
\label{eq:full_posterior}
\end{multline}
where $\epsilon$ is the three-dimensional white-noise field, $\rho_{\rm h}^{\rm data}$ is the observed data in form of halo counts, $\pi(\fnl)$, $\pi(\epsilon)$, and $\pi\left(\left\{b_i \right\}\right)$ are priors, with $\left\{b_i \right\} = \left[\bar{N},b_1,b_2,b_K,b_{\nabla} \right]$ representing the galaxy bias parameters (see Equation \ref{eq:bias_model} for full description). The likelihood, $\pi\left(\rho_{\rm h}^{\rm data}|\epsilon, \fnl,  \left\{b_i 
 \right\} \right)$, is defined by the physical forward model that converts the initial conditions into a prediction of the halo field \citep{jasche_bayesian_2013,andrews_bayesian_2023}.

\subsection{Physics Model}
\label{physics_model}
The forward model transforms the initial conditions into observations such as galaxy data or halo counts. The physical forward model adopted in this paper includes the generation of initial conditions, the application of a linear transfer function, and the simulation of nonlinear structure formation using $N$-body solvers. In our implementation, we assume a standard $\Lambda$CDM cosmology, with modifications to incorporate primordial non-Gaussianity to account for the early-universe perturbations of local PNG.

\subsubsection*{Model of Primordial non-Gaussianity}

In this work, we model primordial non-Gaussianity at the leading order of the local form. The Gaussian field $\phi_{\textrm{G}}$ is obtained by convolving Gaussian white noise $\mvec{\epsilon}$ with a transfer function $T_G(k)$, which ensures that $\phi_{\textrm{G}}$ follows the primordial power spectrum parameterised by $A_{\rm s}$ and $n_{\rm s}$. The covariance of $\phi_{\textrm{G}}$ is:
\begin{equation}
\langle \hat{\phi}_{G,\mvec{a}} \hat{\phi}^{*}_{G,\mvec{b}} \rangle = \delta^K_{\mvec{a},\mvec{b}} T_G(k_\mvec{a})^2,
\end{equation}
where $\hat{\phi}_{G,\mvec{a}}$ are the Fourier modes. The real-space potential is then:
\begin{equation}
\phi_G(\mvec{x}) = \frac{1}{V} \sum_{\mvec{a}} \exp(i \mvec{k}_{\mvec{a}} \cdot \mvec{x}) \hat{\phi}_{G,\mvec{a}} \, ,
\end{equation}
with the non-Gaussian Bardeen potential being defined as:
\begin{equation}
\Phi_{\textrm{NG}}(\mvec{x}) = \phi_{\textrm{G}}(\mvec{x}) +  \fnl \left(\phi_{\textrm{G}}(\mvec{x})^2 - \langle \phi_{\textrm{G}}(\mvec{x})^2 \rangle  \right).
\end{equation}
For more details on the primordial non-Gaussianity model, we refer the reader to \citep[][]{andrews_bayesian_2023,andrews_epng}

\subsubsection*{Forward modelling the matter field}

We obtain the linear matter field, $\delta_{\rm L}$, by rescaling the modes of the Bardeen potential using the matter transfer function, evaluated at $z=9$. We numerically compute the transfer function using the cosmological Boltzmann solver code \texttt{CLASS} \citep[Cosmic Linear Anisotropy Solving System,][]{Lesgourgues_2014}. This field sets the initial conditions for the gravitational structure formation model. 
Unless otherwise stated, we use second-order Lagrangian Perturbation Theory (2LPT) to run the structure formation model. We follow the standard procedure for 2LPT described in \citet{Scoccimarro_1997gr, Crocce_2006, scoccimarro_large-scale_2012}. For more details on the 2LPT structure formation model, we refer to similar work \citep{jasche_bayesian_2013,2015JCAP...01..036J,2016MNRAS.455.3169L,Lavaux2019,2021JCAP...04..033S,Tsaprazi_2022}.
To estimate the late-time, nonlinear density field, we assign the simulation particles to a 3D grid using the Cloud-In-Cell (CIC) mass assignment scheme \citep{hockney_eastwood}.

\subsubsection{Physics-informed priors}
In this work, we adopt a physics-informed prior on the initial white noise field, corresponding to the first prior term described in \citet{2025MNRAS.540..716M}. Specifically, we impose a Gaussian variance prior of the form
\begin{equation}
E_{\mathrm{var}}(x) = \frac{1}{2} \sum_i x_i^2 \, ,
\end{equation}
which enforces zero mean and unit variance of the latent white noise field. This choice reflects the standard assumption of Gaussian initial conditions within the $\Lambda$CDM framework and acts as a regularising term. By constraining the overall variance of the field, the prior suppresses unphysical large-amplitude fluctuations while preserving sufficient freedom for the likelihood to determine the data-constrained modes.

\subsection{Bias model}
\label{sec:bias_model}

In this work, we use a galaxy bias model \citep[][]{desjacques_large-scale_2018} to populate the dark matter field with a halo distribution, defined as \citep{barreira_predictions_2021,cabass_constraints_2022}
\begin{multline}
\label{eq:bias_model}
\rho_{\rm h}(z) = \bar{N} \, \Bigg[ 1 + b_1 \, \delta_{\rm m}(z)  +  \frac{b_2}{2} \delta_{\rm m}^2(z) + b_{K} \, K^2(z) \, + \\b_{\nabla} \nabla^2 \delta_{\rm m}(z) + \, b_\phi \, \fnl \, \phi_{\rm g}(\vec{x}) + \bpd \, \fnl \, \delta_{\rm m}(z) \, \phi_{\rm g} (\vec{x})  \Bigg] \, ,
\end{multline}

where $\bar{N}$ is the mean number of observed halos in the voxels, $b_1$ is the linear bias, $b_2$ is the second-order bias, $b_{K}$ is the tidal field bias, and $K^2(z)$=$\rm{tr}$$\left [ K^2_{ij} (z) \right ]$, where $K_{ij}(z) \equiv \left ( \partial_i \partial_j/\nabla^2 - \delta^{\rm K}_{i,j}/3 \right ) \delta_{\rm m}(z)$ is the long-wavelength tidal field \citep{2021JCAP...10..063L,2021JCAP...08..029B}. The Laplacian bias parameter $b_{\nabla}$ acts on the Laplacian of the density contrast $\nabla^{2}\delta_{\rm m}(z)$, at the assumed resolution of the inference. We note that $\phi(\vec{x})$ is evaluated in Lagrangian space, while $\delta_{\rm m}(z)$ is evaluated in redshift space. However, this distinction has minimal impact at these large scales due to the quasi-local nature of the evolution.

In the model of primordial non-Gaussianity considered, the primordial perturbation introduces a scale-dependent effect on biased tracer populations due to the coupling of short- and long-wavelength modes in a non-zero $\fnl$ universe \citep[][]{dalal_imprints_2008,Slosar_2008,matarrese_effect_2008,Carbone_2008iz,Verde_2009hy}. This coupling leads to a scale-dependent modification in the galaxy bias, which scales as $ \propto k^{-2} $, and is most significant on large scales. In this work, we adopt the universal mass function in a nonzero $\fnl$ context, which allows us to express the bias parameter of the primordial gravitational potential, $ b_{\phi} $, in terms of the linear bias $ b_1 $ as:

\begin{align}
b_{\phi} = 2 \, \delta_{\rm c} \, \left(b_1 - p\right),
\label{eq:bp}
\end{align}
where $ \delta_{\rm c} = 1.686 $ is the spherical critical overdensity in an Einstein--de Sitter universe, and $p$ is fixed to 1.3\footnote{For a motivation of this choice, see the discussion in Section \ref{app:p_sdb}}. Additionally, we include the bias for the cross-field term $ \bpd $, which is modelled as

\begin{align}
\bpd = b_{\phi} - b_1 + 1 + \delta_{\rm c} \, \left[b_2 - \frac{8}{21} \, (b_1 - 1)\right].
\label{eq:bpd}
\end{align}

Precise models for $b_\phi$ and $\bpd$ remain an open challenge in LSS cosmology \citep{biagetti_hunt_2019, achucarro_inflation_2022, sullivan2023learning, 2023arXiv231212405A, 2023arXiv231110088F, 2024arXiv240701391D, 2024arXiv241018039S}. These bias parameters arise from non-linear processes linking the universe's initial conditions to galaxy formation. Although approximate models exist, their discrepancies with simulation results highlight the need for further development to ensure robustness in inferences or formulation of suitable priors \citep{Biagetti_2017,Barreira_2022,barreira2023optimal}. Critically, robust, unbiased, and precise constraints on $\fnl$ from galaxy redshift surveys depend on deploying appropriate models or priors to $b_{\phi}$ and $\bpd$ \citep{Moradinezhad_Dizgah_2021, barreira_local_2021, Lazeyras_2023}. Advancing the theoretical understanding and modelling of these bias parameters will lead to improved and more reliable representations of the underlying physical processes. In this work, we assume the universal mass function to fix the expressions of the bias parameters $b_{\phi}$ and $\bpd$ as above \citep{Barreira_2020b, barreira_local_2021}. With that said, the bias model will be updated as improved models for $b_\phi$ and $\bpd$ become available. A promising approach to understanding the impact of primordial non-Gaussianity on large-scale structure is through assembly bias \citep{2010JCAP...07..013R,2021JCAP...10..063L,2021JCAP...08..029B,2023MNRAS.524..325M,Lazeyras_2023,Lucie_Smith_2023,2023arXiv231110088F,2025arXiv250811798S,2026arXiv260204987P,2026arXiv260212357F}. Previous work has shown that the mass assembly history of halos can significantly influence their clustering properties, and has linked assembly bias to galaxy properties such as colour or emission line luminosity, mitigating the typical degeneracy between $b_{\phi}$ and $\fnl$.

\subsubsection{Gaussian likelihood model for halo formation}

\begin{table*}
    \centering
    \begin{tabular}{p{3.5cm}|p{6cm}|p{6cm}}
    \hline
   Test set-up &    Description  & Main outcomes  \\ \hline \hline
   Fixed initial conditions & Forward model tested with fixed initial \newline conditions. & Recover $\fnlgt$ and estimate of $\sigma_{\rm G}$: Figure \ref{fig:fic_corner}. \\ \hline
   Free initial conditions &  Inferring $\fnl$, initial conditions, and bias parameters in 30 \textit{Quijote}-PNG simulations.   & Estimates of $\fnl$, and cross-correlations with ground truth fields: Figures \ref{fig:highlight}, \ref{fig:mr_cross_pk}.   \\ \hline
   Resolution study \& \newline structure formation model &    Assesses how increasing voxel resolution and choice of structure formation model improve constraints on $\fnl$ using 1LPT, 2LPT, and tCOLA at different scales  & Resolution study, and performance comparison of structure formation models: Figure \ref{fig:resolution_fnl}, \ref{fig:sf_cross_pk}.   \\ \hline
    \end{tabular}
    \caption{Overview of tests performed in this paper}
    \label{tab:overview}
\end{table*}

We assume that the halo counts at the field level follow a Gaussian distribution, and thus the corresponding log-likelihood is given as:
\begin{multline}
    \mathrm{ln} \left [ \pi\left (  \rho_{\rm h}^{\rm data} \, \big| \, \rho_{\rm h}^{\rm pred} \, , \sigma_{\textrm{G}} \right ) \right ] = -N^3 \, \mathrm{ln} \left [\sqrt{2\pi} \, \sigma_{\textrm{G}} \right ] - \\ - \frac{1}{2}\sum_{i = 1}^{N^3} \left ( \frac{\rho_{\textrm{h},j}^{\rm data}-\rho_{\textrm{h},j}^{\rm pred}}{\sigma_{\textrm{G}}} \right ) ^2 \, ,
\label{eq:gaussian_LH}  
\end{multline}
where $\rho_{\textrm{h},j}^{\rm data}$ is the observed halo number count, and $\rho_{\textrm{h},j}^{\rm pred}$ is the model prediction of the physical forward model, given $\epsilon$, $\left\{b_i 
 \right\}$, and $\fnl$; the sum acts over the voxel index $j$ for a total of $N^3$ voxels, which in turn depends on the resolution assumed. We denote $\sigma_{\rm G}$ as the standard deviation of the Gaussian distribution. The Gaussian likelihood and the bias model (Eq. \ref{eq:bias_model}) is valid considering the scales covered in this work, where $k_{\rm max} \approx 0.1\hMpc$, since the density fluctuations are still in the linear or quasi-linear regime at these relatively large scales \citep{desjacques_large-scale_2018,2024arXiv240500635B}.

In this work, $\sigma_{\rm G}$ is inferred and estimated in the case when the initial conditions are fixed to their fiducial values (see Section \ref{fic} for more details). During the inference of the initial conditions, $\sigma_{\rm G}$ remains fixed, to ensure stabilisation of the inference process \citep{2021JCAP...03..058N}. In Appendix \ref{sigmagsigmafnl}, we perform a test run in which we include the joint inferring of $\sigma_{\rm G}$, along with a discussion on inferring $\sigma_{\rm G}$.

\begin{figure*}
\centering
\includegraphics[width=1.75\columnwidth]{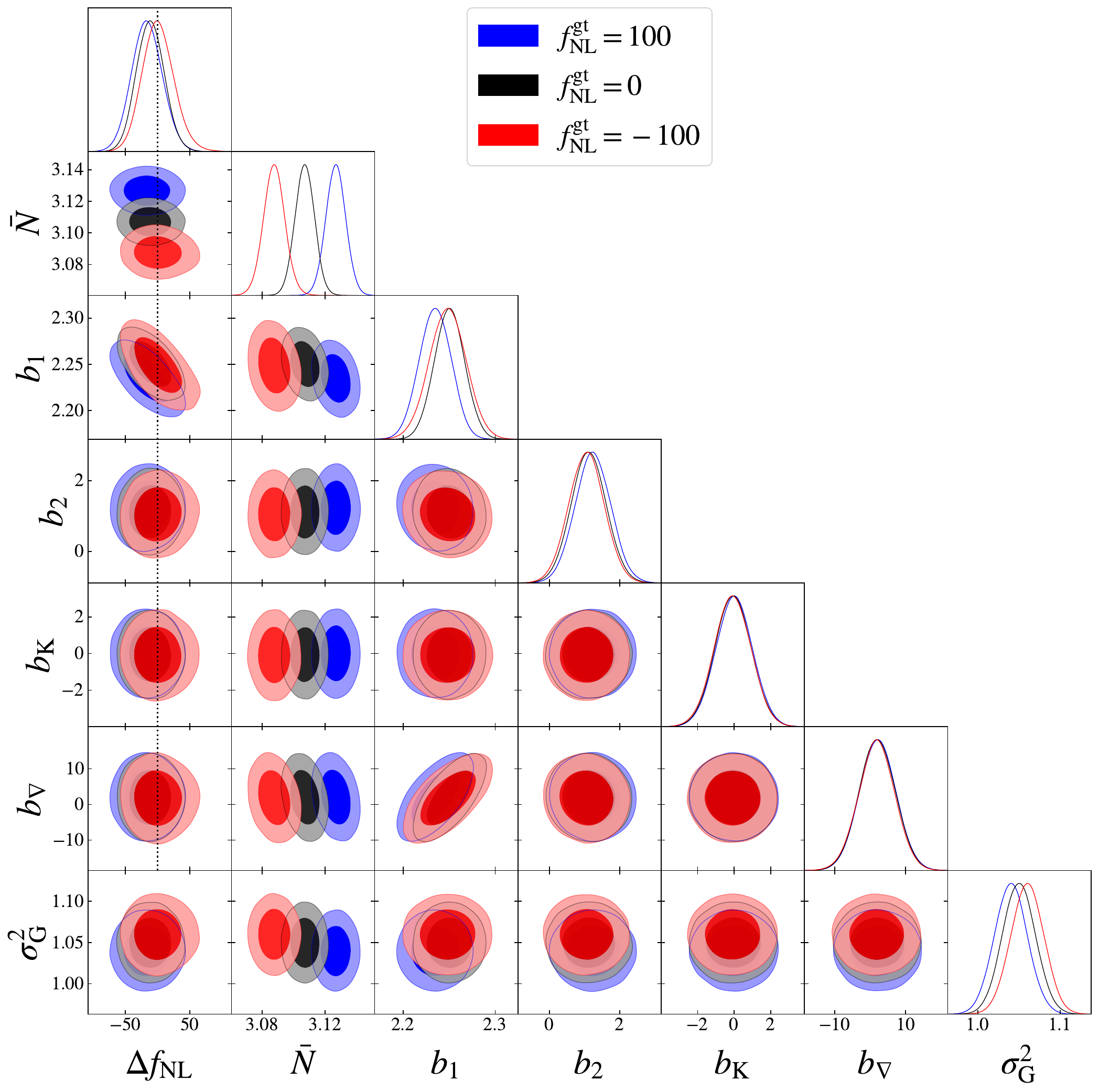}
\caption{Marginalised posteriors of the bias parameters and $\fnl$ for three simulations with fixed initial conditions and $\fnlgt$ values. The results demonstrate that the forward model, including the galaxy bias model, successfully recovers the fiducial $\fnlgt$. These results also provide an estimate for $\sigma_{\rm G}$, which is used in subsequent runs, with the initial conditions fixed to their ground truth values.}
\label{fig:fic_corner}
\end{figure*}

We note that the inferred noise parameter is significantly lower than the Poisson expectation (approximately one-fourth; see the relative values of $\bar{N}$ and $\sigma^2_{\rm G}$ in Figure \ref{fig:fic_corner}), primarily due to the use of the CIC kernel for mass assignment. By distributing each object's mass across adjacent cells, the CIC kernel smooths the field, suppressing small-scale variance and shot noise, and creating a more continuous halo field compared to discrete kernels like the nearest-grid point kernel \cite[][]{2008ApJ...687..738C}, which would result in a "zero-inflated" field with most of the volume empty. This smoothing, combined with the Gaussian noise model's assumption of uniform variance, skews the inferred noise parameter to lower values, as the averaging process is dominated by underdense regions. In short, these deviations from Poisson expectations are thus a consequence of the mass assignment scheme and the noise model, rather than inconsistencies in the framework.

\subsection{Performing the MCMC analyses}
\label{sampling}

\subsubsection*{Sampling algorithm}
A key feature of the previously described forward model is its differentiability. This allows it to be combined with a Hamiltonian Monte Carlo algorithm sampler \citep[][]{1987PhLB..195..216D,betancourt_2017}, which means that the algorithm can be used to include Hamiltonian dynamics to efficiently explore the posterior distribution of Eq. \ref{eq:full_posterior}. On the other hand, $\fnl$ and the nuisance parameters, such as the galaxy bias parameters, are sampled using a slice sampler. The Gibbs block sampling approach is used to update these parameters iteratively \citep{hastings_1970,neal_slice_2003}. For more details, see the corresponding sections in \citep[][]{andrews_bayesian_2023,andrews_epng}.

\subsubsection*{Priors and constraints}
The priors for this analysis are expressed as follows: The initial conditions $\epsilon$ follow a Gaussian distribution $\mathcal{P}(\epsilon) = \mathcal{N}(\bold{0}, \bold{I}^2)$, representing a white-noise field with mean zero and unit variance. We assume the prior on $\fnl$ as $\mathcal{P}(\fnl) \propto \mathcal{N}\left(\fnlgt-\fnl, 100^2 \right) \, \Theta(500 - \fnl) \Theta(\fnl + 500)$, with $\Theta$ being the Heaviside function. The mean number of halos (per voxel), $\bar{N}$, and the linear bias $b_1$ are limited to positive values, $\mathcal{P}(b_1) \propto \theta(b_1)$ and $\mathcal{P}(\bar{N}) \propto \theta(\bar{N})$. Finally, the second-order bias $b_2$, the tidal field bias $b_K$, and the Laplacian bias $b_{\nabla}$ all follow Gaussian priors, expressed as $\mathcal{P}(b_2) = \mathcal{P}(b_K) \propto \mathcal{N}(0, 3^2)$, and $P(b_{\nabla}) \propto \mathcal{N}(0, 5^2)$ with mean 0 and standard deviation 3 or 5. The cosmological parameters used in the analysis are consistent with those used to simulate the data (see Section \ref{data}), and are fixed.

\subsubsection*{Initialisation and convergence}

\begin{figure*}
\centering
\includegraphics[width=2.0\columnwidth]{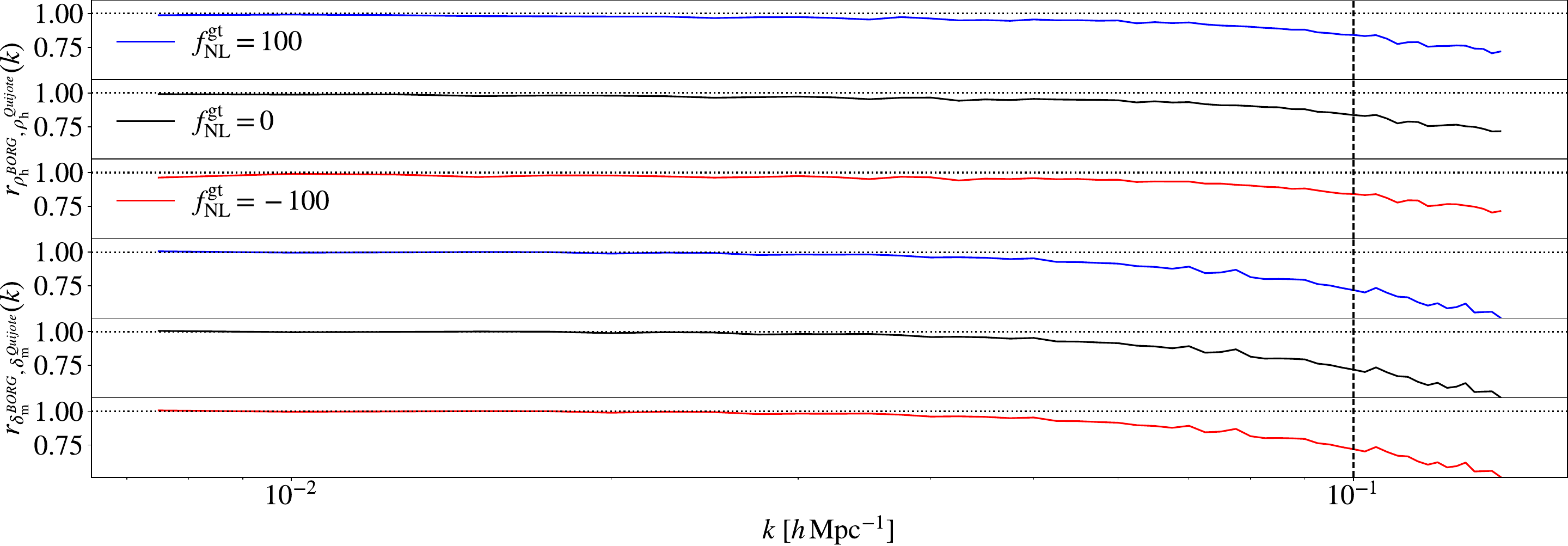}
\caption{The top three panels show the cross-correlation between the inferred and ground-truth halo fields, while the bottom three panels present the cross-correlation between the inferred and ground-truth dark matter density fields. Each cross-spectra is for the same set of fixed initial condition inferences. The decorrelation in the dark matter density cross-correlation arises from model discrepancies in the structure formation model compared to the $N$-body simulations. In the halo field cross-correlation, the model mismatch at smaller scales likely stems from limitations in both the galaxy bias and structure formation models. These results correspond to fixed initial conditions, and analyses and data at $z=0.5$; for results based on sampled initial conditions, see Figure \ref{fig:mr_cross_pk}.}
\label{fig:fic_cross_pk}
\end{figure*}

In this subsection, we outline the process used to initialise and monitor the convergence of the MCMC analyses performed in this paper. The overall goal is to ensure that the chains explore the parameter space efficiently and converge to accurate posterior distributions. We describe the initialisation procedure, including the selection of starting values for the parameters, as well as the steps taken during the warm-up to mitigate biases. In addition, we discuss the methods used to assess convergence, including the calculation of the Effective Sample Size (ESS) and the Gelman–Rubin (G-R) statistic, which help validate the reliability of the MCMC chains. In short, by carefully managing these MCMC runs, we ensure that the final results presented in this paper are robust against initialisation biases or convergence issues.

To ensure that the chains are initialised in an overdispersed state far from the posterior target, the starting values of the density amplitudes $\epsilon$ are drawn randomly and scaled down to $1/10$th of the prior values. The initial values of $\fnl$ are: $\fnl=50 + \fnlgt$. This is done to ensure a reasonable range, based on the expected uncertainty, and to allow the chain to gradually converge toward the true value. The initial values for the galaxy bias parameters are $\bar{N} = 3.1$, $b_1=2.0$, $b_2=b_K=b_{\nabla}=0$. With this initialisation, the MCMC runs can then progress to the warm-up phase.

Once the chain has stabilised and converged, we consider the MCMC analysis to have passed the warm-up phase; the samples that are part of the warm-up phase are discarded from the final analysis to avoid introducing any bias into the final results. During the warm-up phase, we first only sample the phases and number of galaxies, and later we successively sample the other bias parameters. Once these parameters have stabilised, we then sample $\fnl$ as well. This choice is based on the requirement on the algorithm to first converge the large-scale density field with the target posterior, before exploring the posterior of the bias parameters. Allowing the higher-order bias parameters and $\fnl$ to vary freely from the outset could lead to the algorithm trying to overcompensate the low density field, and thus potentially skewing the results.
 
To monitor the convergence of the chains, we compute the ESS of $\fnl$. The ESS measures how many independent samples the chain has effectively produced, allowing us to assess the reliability of the results. For the 30 completed runs, we find the estimated ESS values of $\fnl$ ranging from 1175 to 2731, with a mean and standard deviation of 2133 and 270, respectively.

As an additional check, for the first three simulations, we run an extra MCMC chain for each simulation with different initial parameter values for the MCMC. After discarding the warm-up phase and allowing the chain to converge, we compute the G-R statistic \citep[][]{gr_test_1992}. We aim for convergence by targeting a G-R statistic of $ R \approx 0.01 $. After convergence, the marginal posterior distributions of the chains are consistent. The resulting corner plots with the distributions side by side are shown in Figures \ref{fig:corner_z1z2}, \ref{fig:corner_p1p2}, and \ref{fig:corner_n1n2}. We note that the other chains have a single MCMC chain.

In summary, by consistent and conservative initialisation, implementing phased sampling during warm-up, and assessing convergence through the ESS and G-R statistic, we mitigate potential biases and confirm the statistical robustness of the final posterior distributions.

\subsection{Fisher Forecast Methodology}
\label{sec:fisher_method}

We perform a Fisher matrix forecast to estimate the expected constraints on primordial non-Gaussianity, using power spectrum and bispectrum measurements. Our analysis considers the same cubic survey volume of $1 \, (\mathrm{Gpc}/h)^3$ at redshift $z=0.5$, corresponding to a fundamental frequency of $k_{f} = 2\pi/L\simeq 0.0063 \, h/\mathrm{Mpc}$. For the power spectrum analysis, we use $k_{ \rm \min}^{P} = 0.0063 \, h/\mathrm{Mpc}$, with bin width $\Delta k^{P} = k_{f}$, while for the bispectrum we employ two different binning schemes: one with $k_{\min}^{B} = 0.01 \, h/\mathrm{Mpc}$ and $\Delta k^{B} = 0.01\, h/\mathrm{Mpc}$ and the other with $k_{\min}^{B} = k_{\min}^{P} = 0.0063 \, h/\mathrm{Mpc}$ and $\Delta k^{B} = k_f$. We set the maximum wavenumber to $k_{\max} = 0.1 \, h/\mathrm{Mpc}$ for both the power spectrum and bispectrum analyses. Our Fisher forecast includes redshift-space distortions, and employs the Quijote fiducial cosmology with the following parameters: $\Omega_{\rm m} = 0.3175$, $\Omega_{\rm b} = 0.049$, $H_0 = 67.11$ km~s$^{-1}$~Mpc$^{-1}$, $n_s = 0.9624$, and $\sigma_8 = 0.834$.

The theoretical model for the power spectrum includes both tree-level and one-loop contributions ($P^{1\mathrm{-loop}}$), while the bispectrum model is computed at tree level ($B^{\mathrm{tree}}$). We limit the bispectrum model to tree level in order to match the perturbative order reached by the BORG analysis: since the bias expansion is performed up to second perturbative order, as described in section~\ref{sec:bias_model}, the bispectrum is consistently modelled only at tree level; one-loop contributions would require extending the expansion to higher orders. Both the power spectrum monopole and quadrupole and the bispectrum monopole and quadrupole are included in the likelihood analysis. The assumed number density is the same as the inferred values of the field-level analyses, $\bar{N} = 1.015 \times 10^{-4} \, (h/\mathrm{Mpc})^3$. The fiducial bias parameters are chosen to match the best fit values of the BORG analysis, with values $b_1 = 2.23$, $b_2 = 1.16$, $b_{\nabla^2 \delta} = b_K^2 = 0$, and setting all the third order biases to zero. The priors on the bias parameters are set as follows: $b_1$ and $\bar{N}$ have only positivity constraints, $b_2$ and $b_K^2$ have Gaussian priors with $\pm 3$ around their fiducial values, and $f_{\mathrm{NL}}^{\mathrm{loc}}$ has a Gaussian prior of $\pm 100$. The Fisher matrix is computed with $p_{\mathrm{sdb}} = 1.3$ to compute the non-Gaussian bias, as in eq.~\ref{eq:bp}, which is the same as the main analyses. We use the methodology and the code described in~\citep{Braganca:2023pcp}.

We extract the predicted Fisher information by varying bias parameters and $\fnl$. The expected constraint on the local-type $f_{\mathrm{NL}}$ parameter using power spectrum and bispectrum information ($P^{1\mathrm{-loop}} + B^{\mathrm{tree}}$) is $\sigma(f_{\mathrm{NL}}^{\mathrm{loc}}) = 29$ when $k_{\min}^{B} = 0.01 \, h/\mathrm{Mpc}$ and $\sigma(f_{\mathrm{NL}}^{\mathrm{loc}}) = 27$ when $k_{\min}^{B} = 2\pi/1000 \, h/\mathrm{Mpc}$, compared to $\sigma(f_{\mathrm{NL}}^{\mathrm{loc}}) = 43$ from the power spectrum alone. In standard large-scale structure analyses, most of the constraining power on $f_{\mathrm{NL}}^{\mathrm{loc}}$ arises from the characteristic scale-dependent bias it induces in the galaxy power spectrum on large scales. Because this signature has a very specific scale dependence, it already captures a large fraction of the available information on $f_{\mathrm{NL}}^{\mathrm{loc}}$. The additional information provided by the bispectrum therefore originates primarily from the primordial contribution to the (bias independent) matter bispectrum itself, which provides an additional probe of primordial non-Gaussianity. This explains both the noticeable improvement after including the tree-level bispectrum and the mild dependence, within the range considered, on $k_{\min}^{B}$. We include these results in Figures~\ref{fig:money_plot} and~\ref{fig:highlight}, in comparison with the field-level inference results.

\section{Data}
\label{data}

The \textit{Quijote}-PNG simulation suite is a set of dark matter only $N$-body simulations \citep[][]{2020ApJS..250....2V,Coulton_2023}. Each simulation is set within a cubic volume of $1 \, (h^{-1} \, \text{Gpc})^3$, containing $512^3$ dark matter particles. These simulations are produced using the 2LPTPNG code \citep{Crocce_2006, scoccimarro_large-scale_2012, Coulton_2023} to generate initial conditions at redshift $z = 127$. The evolution of dark-matter particles is subsequently followed to redshift $z = 0$ using the Gadget-III code \citep{2005MNRAS.364.1105S}, which employs a TreePM algorithm. Dark matter halos are identified using the Friends-of-Friends algorithm \citep{1985ApJ...292..371D}, with a linking length parameter $b = 0.2$. The \textit{Quijote}-PNG simulations have enabled numerous studies in the field \citep{Jung_2022,Jung_2023,jung2023quijotepng,2023ApJ...943..178C,2024MNRAS.534.1621C,2023arXiv231110088F,2024arXiv240313985Y,2024arXiv241114377M,2024arXiv240300490J}.

The simulations are based on a fiducial cosmology with parameters $\Omega_{\rm m} = 0.3175$, $\Omega_{\Lambda} = 0.6825$, $n_{\rm s} = 0.9624$, $A_{\rm s} = 1.91\times10^{-9}$ ($\sigma_{8} = 0.834$\footnote{$A_{\rm s}$ was evaluated from $\sigma_8$ at linear scales using \texttt{CLASS} \citep{Lesgourgues_2014}.}), and $H_0 = 67.11$ km~s$^{-1}$~Mpc$^{-1}$. In this study, we used the first 30 simulations, at redshift $z = 0.5$, under different local primordial non-Gaussianity values $\fnl = 0$, $\fnl = +100$, and $\fnl = -100$. The halos considered for analysis are chosen based on a halo mass threshold of $M_{\text{min}} = 3.16 \times 10^{13} h^{-1} \, M_{\odot} = 10^{13.5} h^{-1} \, M_{\odot}$ and upward. The halos are assigned to the data cube using a mass-weighted CIC kernel, and they are analysed in redshift space\footnote{The mapping from real space to redshift space is implemented by displacing halo positions along the line of sight according to their peculiar velocity component in that direction \citep{2023ApJ...943..178C}, and this procedure is applied consistently to both the simulated data and the forward model used in the inference.}.

\section{Results}
\label{results}

\begin{figure*}
	\centering
    \includegraphics[width=2.0\columnwidth]{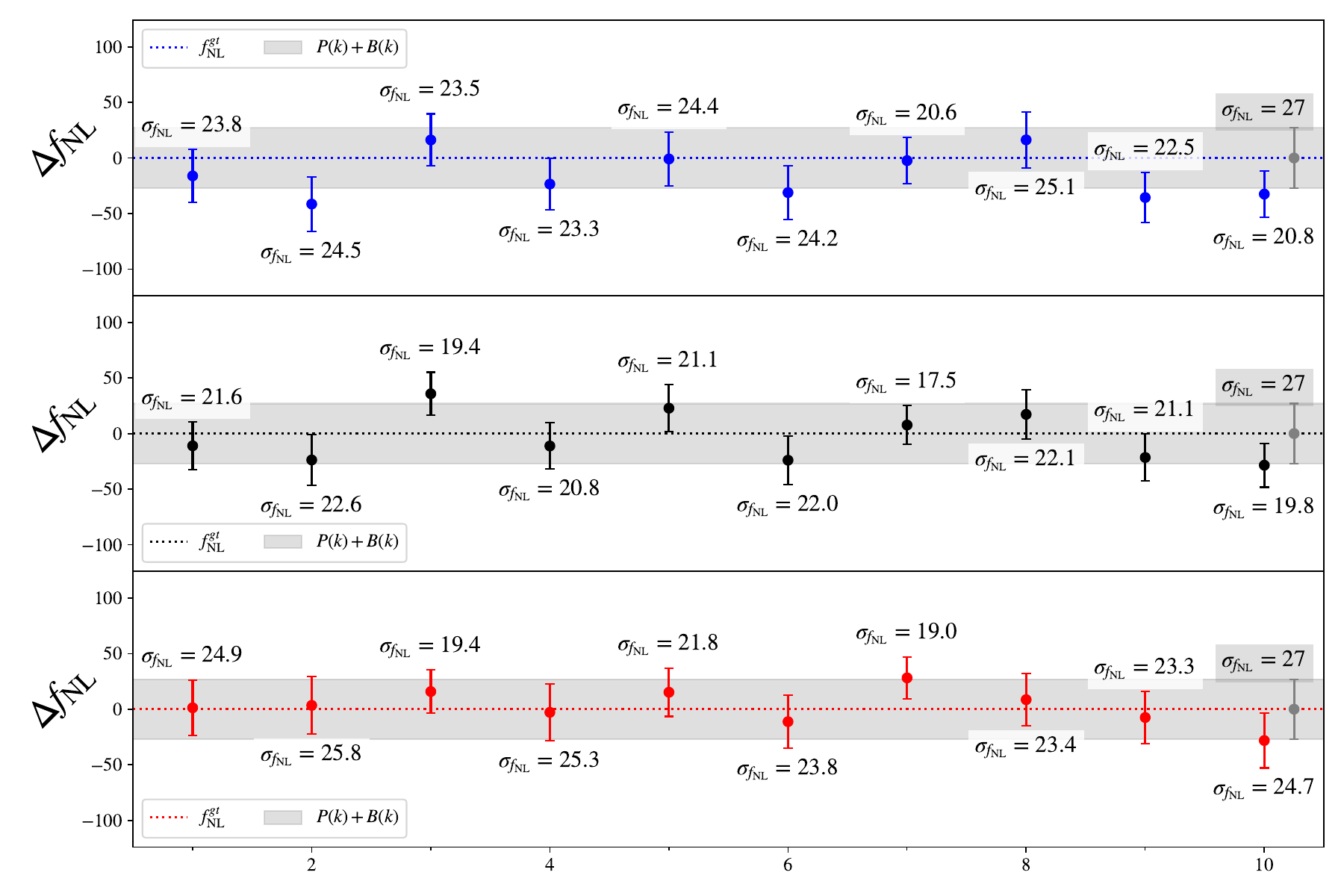}
    \caption{Comparison of inferred $f_{\rm NL}$ values over 30 simulations, split up between positive $\fnlgt$ (top row), null $\fnlgt$ (middle row), and negative $\fnlgt$ (bottom row). The error bars represent the uncertainty $\sigmafnl$, and the horizontal dashed lines represent the ground-truth $\fnlgt$ value. The results showcase the capability of the field-level inference algorithm to recover the $\fnlgt$ while also marginalising over initial conditions and bias parameters (with $b_{\phi}$ and $\bpd$ fixed as in Eqs. \ref{eq:bp} and \ref{eq:bpd}).}
	\label{fig:highlight}
\end{figure*}

\begin{figure*}
	\centering
    \includegraphics[width=1.6\columnwidth]{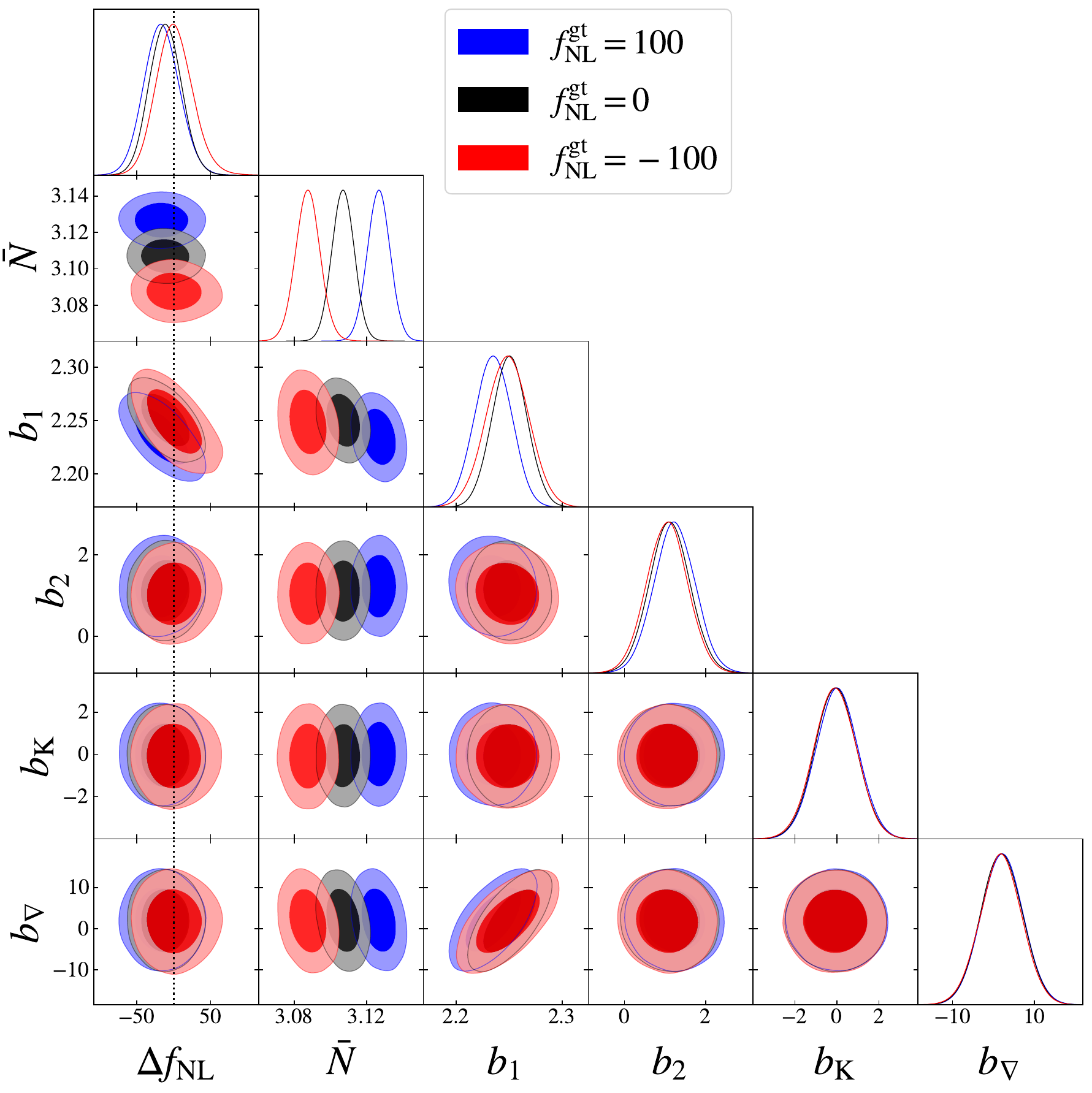}
	\caption{Similar to Figure \ref{fig:fic_corner}, but with initial conditions being sampled. $\sigma_{\rm G}^2$ is fixed to 1.2. For this simulation and the given resolution, our method is able to recover the $\fnlgt$. Notice the degeneracies between $\fnl$ and $b_1$, as well as between $b_1$ and $b_{\nabla}$. The inferred $\fnl$ have been shifted to $0$ for visualisation.}
	\label{fig:mr_corner}
\end{figure*}

In this section, we describe each test configuration and present the resulting outcome. Table \ref{tab:overview} offers an overview of the tested configurations. In Section \ref{fic}, we discuss the case of fixed initial conditions set up, where we test the forward model and the adopted bias model. We then proceed with the joint sampling of the initial conditions in Section \ref{std}, and provide a series of validation tests of the inference. In Section \ref{resolution_sf_study}, we test the sensitivity of $\fnl$ to small-scale physics. 

\subsection{Validation of the forward model using fixed initial conditions}
\label{fic}
The aim of this section is to provide a baseline to assess the model's ability to recover the ground truth $\fnlgt$ and the underlying matter density field. By comparing the model predictions with the data, we quantify and measure the mismatch of the model with the parameter $\sigma_{\rm G}$. To evaluate the performance of our forward model, we begin by testing it in an idealised setting where the fiducial initial conditions $\epsilon$ are known in terms of phases and amplitudes. The goal is to determine whether the model can accurately recover $ \fnlgt $, and reproduce halo statistics which correlate with ground truth fields. In short, this test step is critical to ensuring the reliability and validity of the forward model in subsequent analyses where initial conditions are inferred rather than fixed.

In this test, to generate the initial conditions, we use the 2LPTIC code \citep{scoccimarro_large-scale_2012,Coulton_2023} to compute the ground truth potential field, $\phi_{\rm g}^{\rm gt}$, which corresponds to the observed halo density, $\rho_{\rm h}^{\rm data}$. We then use this field as input to the \borg{} forward model (from the point of the perturbation of $\fnl$ and onwards). With this setup, we sample only the bias parameters and $\fnl$, along with the noise level parameter $\sigma_{\rm G}$ from a Gaussian likelihood. This approach allows us to, firstly, evaluate how well the 2LPT structure formation model, at the given resolution, reproduces the full $N$-body simulation. Secondly, we test whether the bias model can accurately recover the halo statistics. The inference is performed on a grid of $N=32^3$ voxels, resulting in a voxel size of $L_{\rm voxel}=1000\Mpch/32=31.25\Mpch$, which corresponds to $k_{\rm max}\approx0.1\hMpc$.

The results of the inferred bias parameters and $\fnl$ are displayed in the corner plots of Figure \ref{fig:fic_corner}, after discarding the samples from the warm-up period (on the order of $100$ samples). From these results, we can see that the forward model successfully recovers the ground truth values of $\fnlgt$, for each of the three cases, indicating that the model is sufficiently complex to model the halo statistics and distinguish the effect of $\fnl$ in the data. There is a small expected degeneracy between $\fnl$ and the linear bias $b_1$, due to the adopted model of scale-dependent bias (Eqs. \ref{eq:bp} and \ref{eq:bpd}).

To visualise the 2-pt correlation, in Figure \ref{fig:fic_cross_pk} we show cross-correlation spectra, $r(k)$, which quantifies the degree of alignment of the ground truth fields $\delta_{\rm m}^{\rm gt}$ and $\rho^{\rm data}_{\rm h}$ with their inferred counterparts. The cross-spectra demonstrate a high level of correlation between the halo field and the matter field at large scales, where the bias model and structure formation model are expected to perform most accurately. At smaller scales, we observe a gradual drop in the cross-correlation, which is expected due to the increasing nonlinearity of the small-scale structure formation processes in the full $N$-body simulation, which is not sufficiently captured by the forward model. We emphasise that these results reflect the accuracy of the physical forward model in the idealised case of fixed initial conditions, rather than the performance of \borg{}'s sampling procedure or its ability to infer initial conditions. For insights into the latter, see the results in Figure \ref{fig:mr_cross_pk} for the case when initial conditions are inferred.

In addition to recovering $\fnl$, we also obtain an estimate for the noise level $\sigma_{\rm G}$, which represents the uncertainty or model mismatch in the inference process. This value of $\sigmafnl$ will be used in the analyses covered in the next section when inferring the initial conditions $\epsilon$, allowing us to fix the noise level and avoid destabilising the inference \citep[such as "sigma collapse" or long correlation lengths, as reported in][]{2021JCAP...03..058N}. For other works dealing with this problem through informative priors, see \citet{2024arXiv240303220N,2024arXiv240502252B}.

This initial validation indicates that the forward model is able to recover the underlying density field and $\fnlgt$, confirming the suitability of the forward model for the simulated data. Next, we run the inference while also sampling the initial conditions.

\subsection{Main analysis: Joint inference of $\fnl$, bias parameters, and initial conditions}
\label{std}

\begin{figure}
	\centering
    \includegraphics[width=1.0\columnwidth]{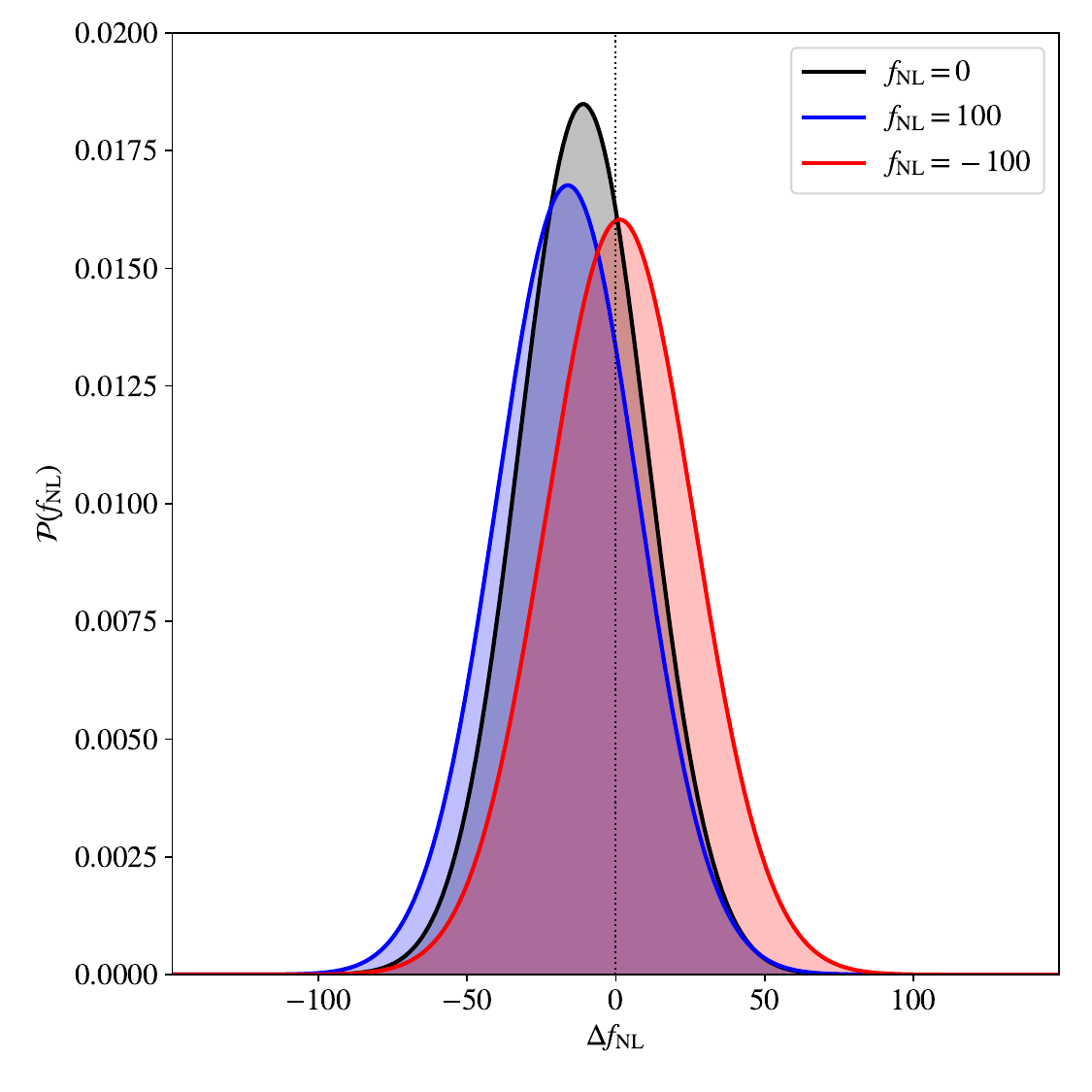}
	\caption{
        The inferred 1D $\fnl$ posterior distributions for first three simulations, with $\fnlgt=100$ in blue, $\fnlgt=0$ in black, and $\fnlgt=-100$ in red. Each posterior has been shifted to be centred at $\fnl=0$.
    }
	\label{fig:fnl_pdfs}
\end{figure}

In this section, with the validated forward model in hand, we apply our method to a subset of simulations from the \textit{Quijote}-PNG suite, with the goal of inferring and recovering the ground truth values of $\fnl$, initial conditions, and bias parameters. Our approach aims to provide tighter constraints on $\fnl$ compared to traditional estimators based on power spectra and bispectra, while simultaneously reconstructing the initial conditions. By incorporating field-level information, we aim to make a significant step in the direction of more precise cosmological analyses. For simplicity, we fix $b_{\phi}$ and $\bpd$ to the expressions described in Eqs. \ref{eq:bp} and \ref{eq:bpd}.

To properly test the algorithm under realistic conditions, our analysis configuration includes sampling the initial conditions, bias parameters, and $\fnl$, and follows the procedure described in Section \ref{sampling}. The analysis is applied to 30 simulations across the three ground truth values of $\fnlgt$ ($0$, $100$, and $-100$). The noise parameter is fixed to $\sigma^2_{\rm G}=1.2$, which is the upper range of the posterior as inferred by the set of fixed initial conditions (Section \ref{fic}). This upper choice is motivated by the need to provide additional robustness to the mismatch of the model. In Appendix \ref{sigmagsigmafnl}, we also perform a test run where $\sigma_{\rm G}$ is also inferred; we also test how the inferred $\sigmafnl$ depends on the assumed $\sigma_{\rm G}$.

As a first point, we present the inferred $\fnl$ values for the 30 runs. An overview of recovered $\fnl$ is provided in Figure \ref{fig:highlight}, with ground truth $\fnl$ within the $1\sigmafnl$ range in approximately 68\% of the cases. This highlights the consistency of the method in accurately recovering the ground truth value of $\fnlgt$, even when marginalising over the bias parameters and initial conditions, with $b_{\phi}$ and $\bpd$ fixed to the universality relation. The grey band in the figure represents the Fisher-forecasted constraints achievable by the combined use of power spectra and bispectra estimators.

We note the asymmetric uncertainties in the inferred values of $\fnl$; larger for positive $\fnl$ and smaller for negative $\fnl$. The probability distribution of the large-scale power spectrum can be modelled as following an inverse gamma distribution \citep{Wandelt_2004,Jasche_2013lwa,jasche2017}, which naturally has long tails and is non-symmetric. Since the scale-dependent bias effect is most pronounced on the largest scales, increasing the amplitude of the power spectrum will lead to increased scatter in inferred $\fnl$ values. Vice versa, for negative $\fnl$, the suppression of clustering reduces sensitivity to cosmic variance and limits the influence of long-tailed uncertainties, resulting in tighter constraints. In short, the observed asymmetry in the uncertainties is due to the interplay between the scale-dependent bias effect and the statistical properties of the large-scale power spectrum.

We visualise the marginal posterior distributions of the bias parameters and $\fnl$, in the first runs, in the corner plot of Figure \ref{fig:mr_corner}. This figure illustrates the joint distributions and highlights the degeneracies between certain parameters, particularly between $b_1$ (the linear bias) and $\fnl$, which is an expected result due to the adopted model of scale-dependent bias. Despite these degeneracies, the model successfully recovers the ground truth $\fnlgt$, demonstrating that the method is robust in jointly constraining both bias parameters and $\fnl$. We visualise the 1-D posterior distribution of $\fnl$ for the three first-most simulations in Figure \ref{fig:fnl_pdfs}.

\begin{figure*}
	\centering
    \includegraphics[width=1.95\columnwidth]{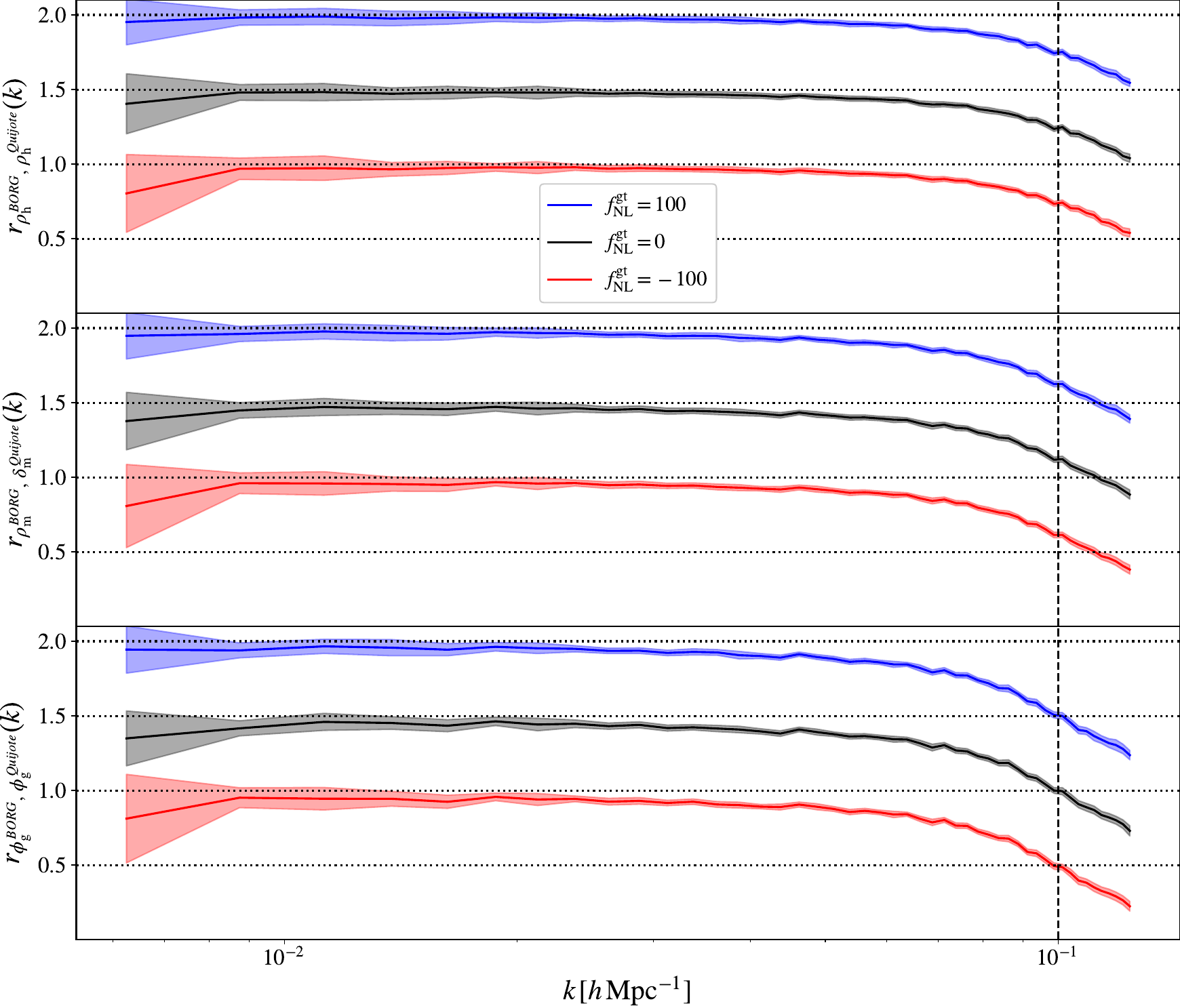}
	\caption{Cross-correlation power spectra between the ensemble of inferred and fiducial halo field (top), dark matter density field (mid), and the gravitational potential (bottom), for $\fnlgt = 100$ (blue), $\fnlgt = 0$ (black), $\fnlgt = -100$ (red), for the sampled initial conditions set-up. The results for $\fnlgt=100$ and $\fnlgt=0$ have been shifted by $1.0$ and $0.5$, respectively, for clarity. Similar to Figure \ref{fig:fic_cross_pk}, in the cross-correlation with the underlying dark matter density field, the decorrelation at smaller scales is due to the model mismatch in the structure formation. The cross-correlation with the gravitational potential describes how well the method can recover the ground truth initial conditions. The analyses and data are at $z=0.5$}
	\label{fig:mr_cross_pk}
\end{figure*}

To evaluate the precision of initial condition inference in 2-pt statistics, we compute the cross-correlation power spectrum $r(k)$ (Eq. \ref{eq:cross_correlation}) between the inferred and ground-truth fields. Figure \ref{fig:mr_cross_pk} displays the cross-correlation between $\phi_{\rm g}$ (gravitational potential), $\delta_{\rm m}$ (matter density field), and $\rho_{\rm h}$ (halo density field). The results show a significant degree of correlation at large scales (small $k$), confirming the model's ability to recover the large-scale structure. At smaller scales, we observe a drop in correlation, likely due to the increasing nonlinearity and complexity of small-scale structures that the forward model does not fully capture.

\begin{figure*}
	\centering
    \includegraphics[width=2.0\columnwidth]{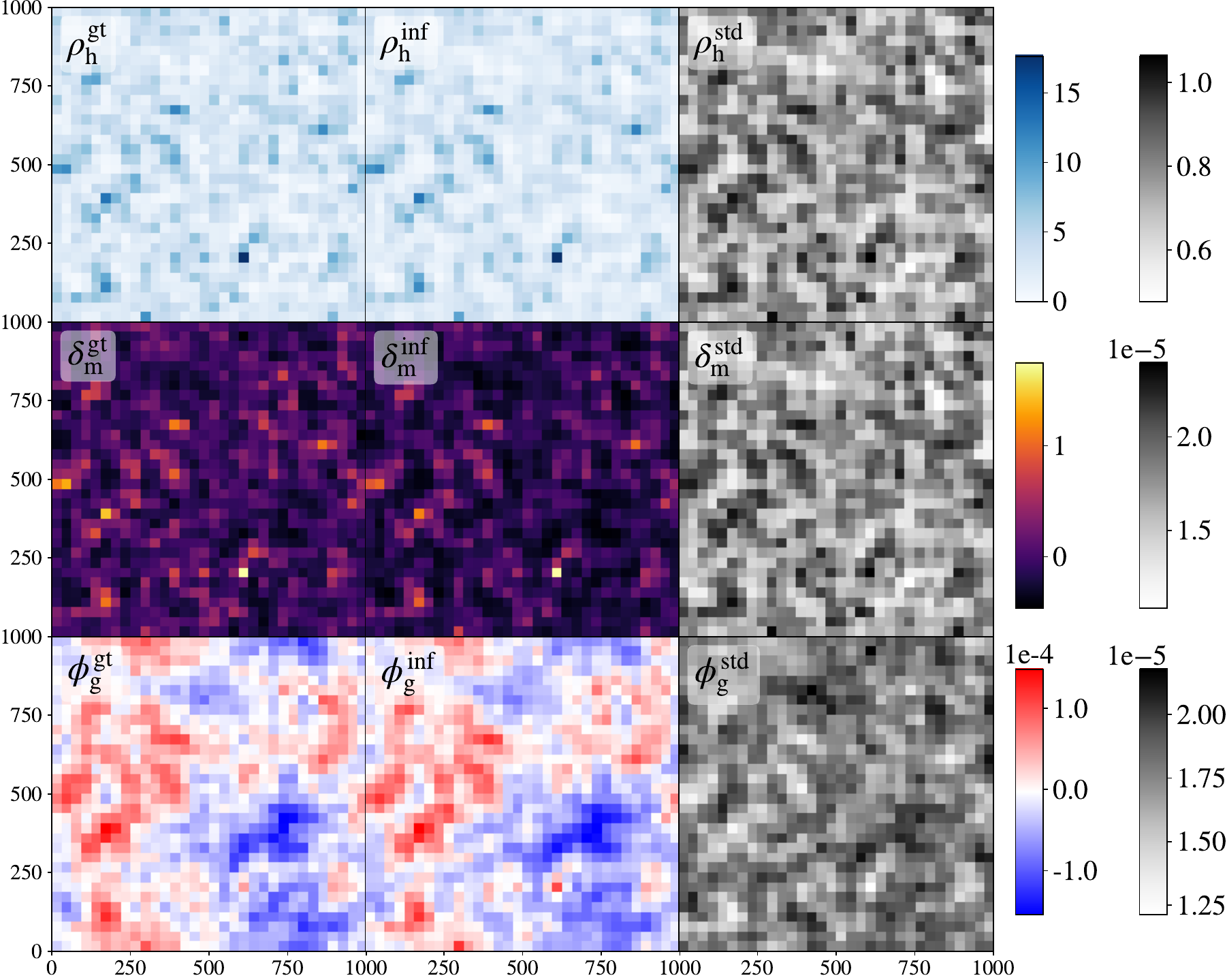}
	\caption{Slices through the halo fields $\rho_{\rm h}$, density fields $\delta_{\rm m}$, and gravitational potentials $\phi_{\rm g}$, for the ground truth (gt), inferred mean (inf), and uncertainty (std). In each row, the left colour bar corresponds to the two left subplots, while the right colour bar corresponds to the right subplot. The voxel sizes are $31.25 \, \Mpch$. The distances shown are in $\Mpch$. The computed Pearson correlation coefficients are 0.970 $\left (\rho^{\rm data}_{\rm h} \, \rm{and} \, \rho_{\rm h}^{\rm inf} \right )$, 0.905 $\left (\delta^{\rm gt}_{\rm m} \, \rm{and} \, \delta_{\rm m}^{\rm inf} \right )$, and 0.944 $\left (\phi^{\rm gt}_{\rm g} \, \rm{and} \, \phi_{\rm g}^{\rm inf} \right )$.}
    \label{fig:3_by_3}
\end{figure*}

To visually inspect the inference, we provide field-level comparisons between ground truth and inferred fields. Figure \ref{fig:3_by_3} shows 2D slices through the 3D cubes of $\phi_{\rm g}$, $\delta_{\rm m}$, and $\rho_{\rm h}$. Visual comparisons show that the inferred fields align closely with the ground truth, particularly on large scales. However, as expected, the inferred fields appear smoother because of the averaging effect of the MCMC sampling process and limitations of the forward model in capturing extreme small-scale features. We also include the 1-pt correlation between the inferred mean and the ground truth fields, in the form of Pearson's correlation coefficients \citep{Pearson1895}.

\begin{figure}
	\centering
    \includegraphics[width=1.0\columnwidth]{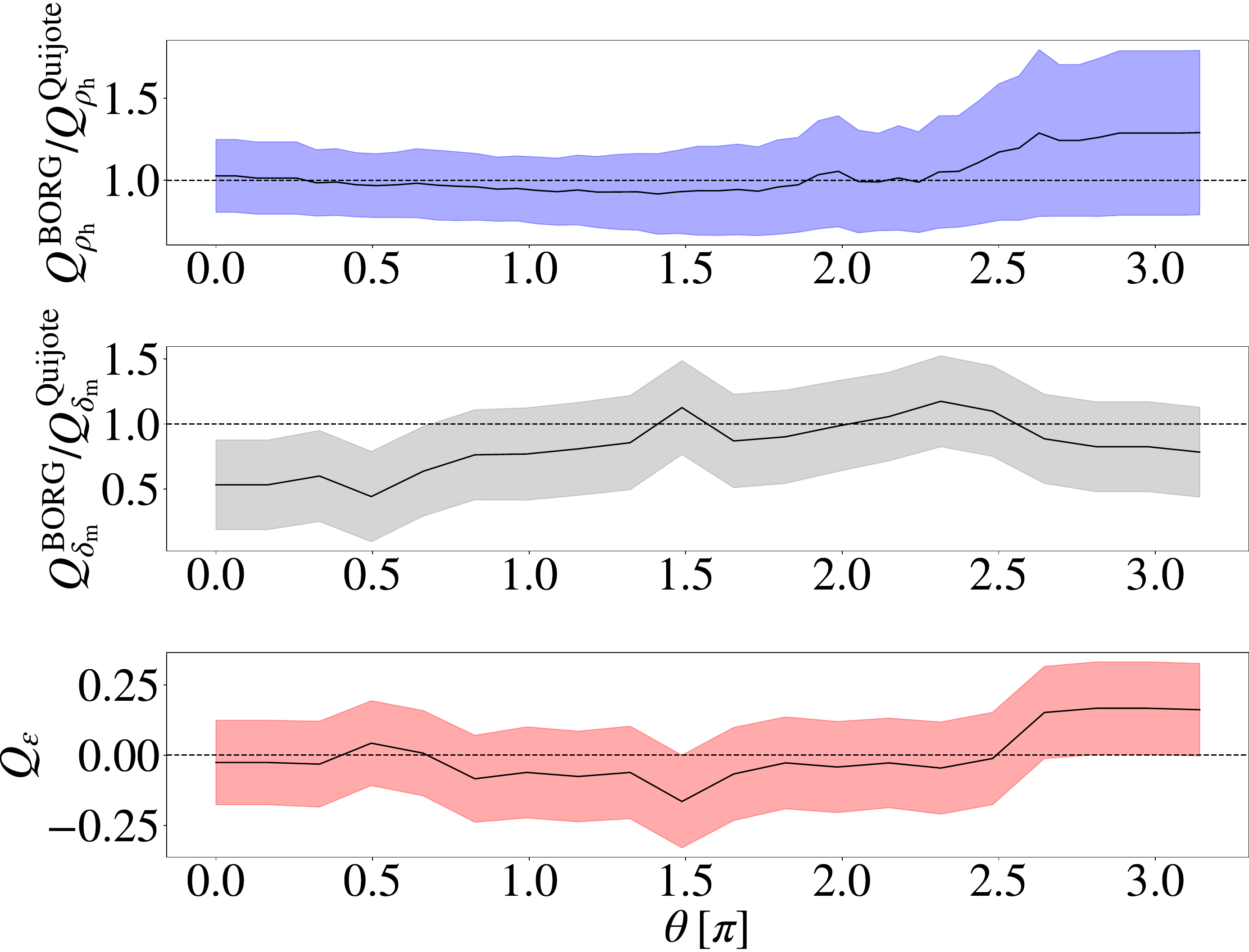}
	\caption{Comparison of (reduced) bispectra derived from \borg{}-predicted fields and the \textit{Quijote} simulations (the ground truth). Top panel: Ratio of the bispectrum for the predicted halo field, $Q_{\rho}(k)$, to that of the \textit{Quijote} halo field. Middle panel: Ratio of the bispectrum for the predicted density field, $Q_{\delta}(k)$, to that of the \textit{Quijote} density field. Bottom panel: The bispectrum for the inferred initial conditions, $Q_{\epsilon}(k)$, relative the expected reference value ($0$). Mean values are shown by the solid lines, with $1\sigma$ uncertainty highlighted by the shaded region. Horizontal dashed lines represent the baseline values for each ratio (unity for the halo and density fields, zero for the initial conditions).}
    \label{fig:bk_ratios}
\end{figure}

We also include evaluations of the recovered 3-pt statistics of the various fields. In Figure \ref{fig:bk_ratios}, we present the bispectrum ratios comparing the inferred bispectra with the corresponding ground truth fields. The reduced bispectrum is defined as the bispectrum $B(k_1, k_2, k_3)$ normalised by the power spectra terms:
\begin{align}
    Q(k_1, k_2, k_3) = \frac{B(k_1, k_2, k_3)}{P(k_1)P(k_2) + P(k_2)P(k_3) + P(k_3)P(k_1)}.
\end{align}
Here, $ k_1 $, $ k_2 $, and $ k_3 $ form a closed triangle in Fourier space. For the range of scales considered in the reduced bispectrum, $k_{\text{min}} = 0.006 \, \hMpc$ and $k_{\text{max}} = 0.1 \, \hMpc$, the angle $\theta$ describes the relative amplitude in each triangle configuration. As $\theta \rightarrow \pi$, the bispectra ratios, especially for $\rho_{\rm h}$ and $\delta_{\rm m}$, align more closely with the reference value. This is expected, since at such configurations the bispectra triangles are closer to the "squeezed" limit, which is more sensitive to large-scale modes. The reduced bispectra are computed using \texttt{Pylians3}\footnote{\url{https://pylians3.readthedocs.io/}}.

In summary, in this section, we describe the application of field-level inference algorithm to simulations from the \textit{Quijote}-PNG suite. We show that we can make statistically robust inferences of $\fnl$, the initial conditions, and the bias parameters. By incorporating field-level information, our approach yields tighter constraints on $\fnl$, similar to the results reported in \citet{2024arXiv240303220N} for $\sigma_8$.
Our reconstructed initial conditions and dark matter density fields display consistency with the ground truth fields, in terms of 1-pt, 2-pt, and 3-pt statistics. Our findings demonstrate that field-level inference provides an alternative for constraining $\fnl$.

Next, we conduct additional runs where we test the response in $\fnl$ constraints to access to small-scale information.

\subsection{Investigating the sensitivity to small-scale information}
\label{resolution_sf_study}

In this study, we investigate how the availability of small-scale information in the forward model impacts the precision of the constraints on $\fnl$. Specifically, we focus on two aspects: (1) a resolution study that explores how increasing resolution affects the inference and constraints, and (2) an analysis of how the choice of structure formation model affects the constraints. Using the same data (simulation index \#1 at $\fnl = 0$), we run a total of five additional inference chains that differ only by their adopted forward model. The aim is to assess how much small-scale physics contributes to the constraints on $\fnl$, for a given realistic dataset.

In the first three inference runs, we use a 1LPT forward model and vary the assumed resolution: $N=8,16,$ or $32$, resulting in voxel sizes $L_{\rm voxel}=125,62.5,$ and $31.25 \, \Mpch$ ($k_{\rm max} \approx 0.025, 0.05,$ and $0.1$ \hMpc), respectively. For all runs, the largest scale is defined by the simulation cube, $k_{\rm min} = 2\pi/L_{\rm box}$. Secondly, for the fixed resolution of $L_{\rm voxel}=31.25 \, \Mpch$, we run three separate chains, with the structure formation models (i) 1LPT, (ii) 2LPT, and (iii) temporal Comoving Lagrangian Acceleration \citep[tCOLA;][]{tassev_solving_2013}. 

The tCOLA solver is initialised from $z_{\rm start}=127$, with $64^3$ particles and $20$ timesteps. The gravitational potential is evaluated with $128^3$ voxels. For a short description of the $N$-body solver tCOLA, we refer the reader to Appendix \ref{tcola_sect}. For a more thorough description of tCOLA, we refer the reader to the relevant literature \citep{tassev_solving_2013}.

Figure \ref{fig:resolution_fnl} presents a summary of the inference results for the three resolutions and the three different structure formation models. As resolution increases, the forward model has access to information on smaller scales, leading to a noticeable improvement in the constraints on $\fnl$. This finding is echoed in \citet[][]{andrews_bayesian_2023,2024arXiv241018039S,andrews_epng}. However, we find a minor impact from the use of higher-fidelity structure formation models on the constraints of $\fnl$, at the assumed resolution. The chains use the same initialisations as the main runs.

\begin{figure*}
	\centering
    \includegraphics[width=2.0\columnwidth]{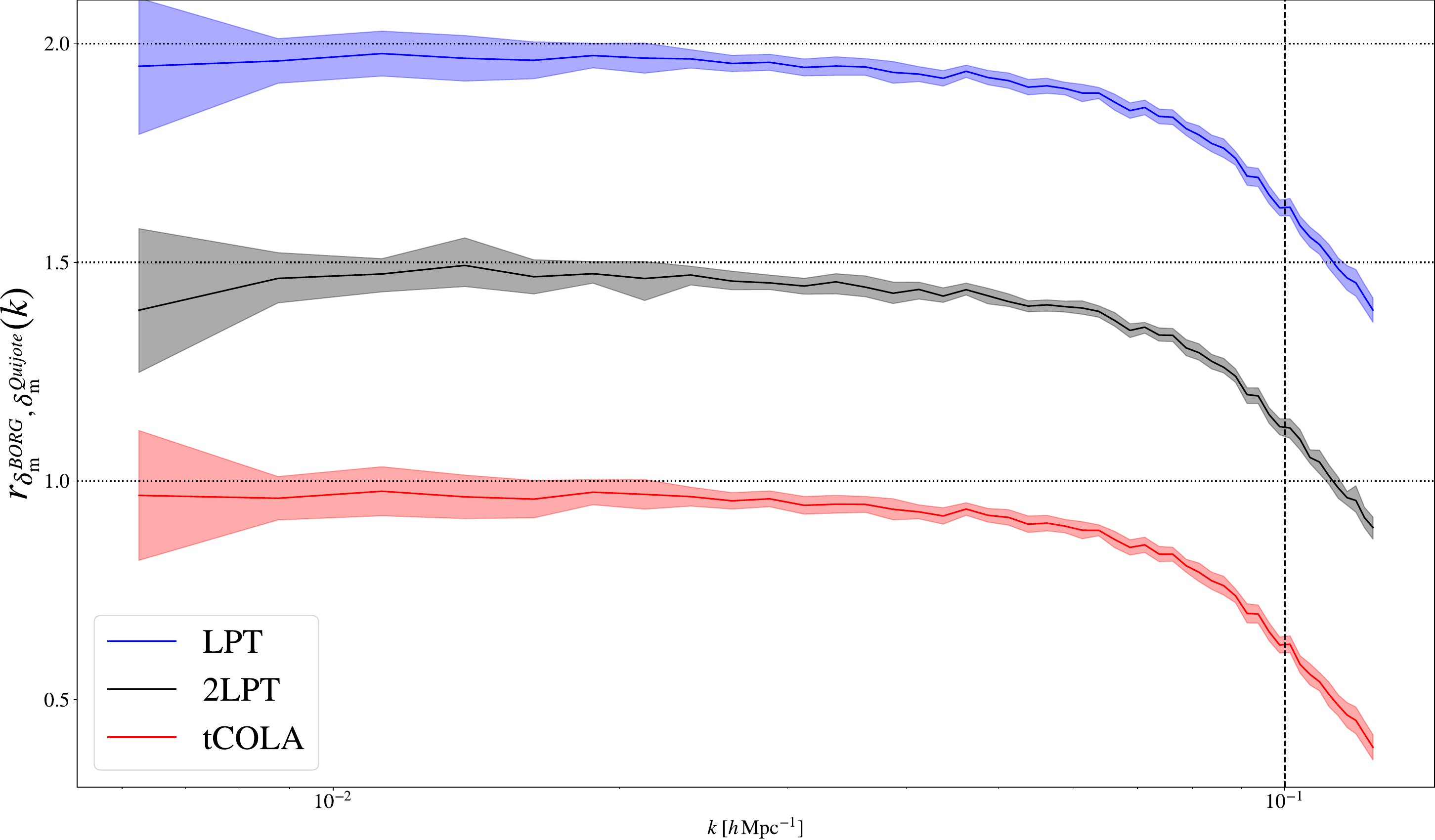}
	\caption{As similar to Figure \ref{fig:mr_cross_pk}, but only $\delta_{\rm m}$ for the structure-formation case study. The results indicate that in the scales considered, higher-order corrections and time-stepping schemes have a negligible impact on the cross-correlation. In the upper panel, the cross-spectrum between the inferred and ground truth dark matter fields, including uncertainty estimates, are shown, where the 1LPT and 2LPT runs have their results slightly shifted upwards. The lower panel shows the same results as the upper panel, but the mean and standard deviation are unshifted.}
	\label{fig:sf_cross_pk}
\end{figure*}

Figure \ref{fig:sf_cross_pk} illustrates the cross-correlation between inferred and ground truth $\delta_{\rm m}^{\rm gt}$ for each run. By comparing the different forward models with varying degrees of small-scale modelling, the results suggest that more accurate small-scale physics modelling does not lead to an increase in cross-correlation at smaller scales with the ground truth $\delta_{\rm m}^{\rm gt}$, given the scale of the inference and the bias model adopted, which is still within the linear regime (voxel size of $31.25 \, \Mpch$). For inferences at higher resolutions than those considered in this paper, this result is not expected to hold, since simpler forward models fail to capture nonlinear effects, leading to biases in inferred parameters $\fnl$ or bias parameters\footnote{We note that the inferred noise parameter may vary across the three runs, but as it is fixed at $\sigma^2_{\rm G} = 1.2$ in this test, this remains unconfirmed.}.

\section{Summary and Conclusion}
\label{conclusions}

The physical origin of primordial fluctuations remains one of the central open questions in the standard $\Lambda$CDM cosmological model, with cosmic inflation offering a compelling framework for their generation. A key observable tied to inflationary physics is local primordial non-Gaussianity, parameterised by $\fnl$, which serves as a probe of the number of active fields during inflation. Upcoming large-scale galaxy surveys, such as those planned by next-generation observatories, promise to place stringent constraints on $\fnl$. However, achieving these goals requires advanced data analysis techniques capable of fully exploiting the statistical power of these data sets. Field-level inference is a cutting-edge approach that allows the reconstruction of initial conditions and precise parameter estimation, making it a promising tool to investigate primordial non-Gaussianity with unprecedented sensitivity. In this context, our study aims to evaluate the performance of a field-level inference framework when applied to simulated halo catalogues, explore its ability to recover $\fnl$ while accounting for model uncertainties and nuisance parameters, and compare the resulting constraints with those achievable with combined power spectrum and bispectrum estimators.

In this work, we have presented an analysis of the performance of our field-level inference framework when applied to the halo catalogues of the \textit{Quijote}-PNG simulation suite. Our main goal is to evaluate the method's capability to recover the local primordial non-Gaussianity parameter $\fnl$, while marginalising over the initial conditions and nuisance parameters. By running the inference in various configurations, we test the robustness and performance of our method. This analysis represents a major milestone, as it serves as a necessary step that our framework must pass to ensure its readiness for application to next-generation data to constrain $\fnl$.

As a first step, we test our approach in an idealised scenario where the ground-truth initial conditions are known. This demonstrated that our forward model is sufficiently complex to recover the ground truth value $\fnlgt$. Moreover, in this setting, we obtain an estimate of the noise parameter of the Gaussian likelihood $\sigma_{\rm G}$ to quantify the model mismatch. The results indicate that the forward model successfully captured the ground truth values, as seen in Figure \ref{fig:fic_corner}.

Building on this, we tested our algorithm on a subset of 30 simulations, where we sampled the initial conditions, the galaxy bias parameters, and $\fnl$. Our findings show that the method recovers consistently $\fnlgt$ within the expected confidence interval throughout the simulations. Compared to traditional estimators that rely on a combination of power spectra and bispectra, our approach provides an improvement in constraints (Figures \ref{fig:money_plot} and \ref{fig:highlight}). 
We demonstrate the consistency of our inferred fields by showcasing the correlation in the 1-point, 2-point, and 3-point statistics. The results show strong correlations between the inferred and ground-truth density fields on large scales, whereas decorrelation occurs at smaller scales, showcasing the impact of the model mismatch in the structure formation model.

We also investigate the sensitivity of the measured $\fnl$ value to small-scale physics. In the first part, we demonstrate that increasing the resolution of the inference leads to increased constraints on $\fnl$. These promising results align with the findings of \citep[][]{andrews_bayesian_2023,andrews_epng}, which account for realistic survey effects and deploy a simplified treatment of redshift-space distortions. In the second part, we tested the impact of different structure formation models -- at the given resolution ($k_{\rm max} \approx 0.1 \hMpc$), there is no substantial impact on the constraints on $\fnl$ or the cross-correlation with the underlying dark matter field as a response to the improved fidelity of the structure formation model.

In conclusion, this study highlights the potential of field-level inference to constrain primordial non-Gaussianity in halo catalogues. This study demonstrates the power of field-level inference as a novel tool for cosmological data analysis, capable of extracting detailed information about primordial non-Gaussianity and cosmological formation history, as well as their readiness for application to forthcoming large-scale structure surveys. In light of these findings, our results underscore the importance of deploying statistical forward-modelling techniques to extract detailed information from galaxy fields. By applying these novel methods to upcoming datasets, future and ongoing cosmological experiments can perform more sensitive investigations into the early universe and go beyond what would be possible with traditional analysis techniques.

\subsection{Future Work}

\begin{figure}
	\centering
    \includegraphics[width=1\columnwidth]{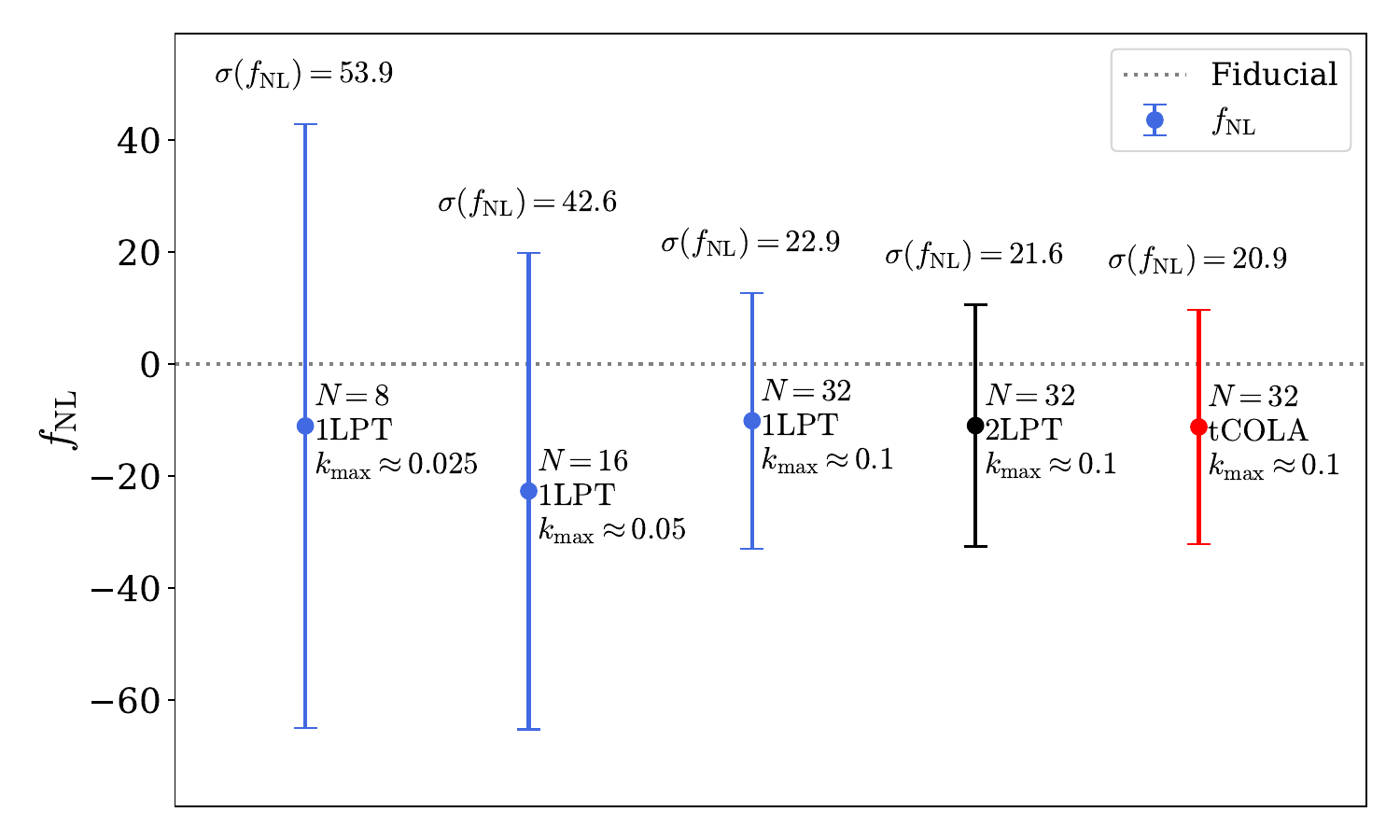}
	\caption{The inferred $\fnl$ values of the five runs for the comparison of small-scale information. The $k_{\rm max}$ values are given in $\hMpc$. Over the resolution increase (the three left-hand points), the recovered $\fnl$ constraints improve. Over the change in the structure formation models (the three right-hand points), there is marginal improvement in the $\fnl$ constraints, suggesting that the small-scale information is slightly relevant for the inference at these scales, and minor improvements can be made as the fidelity of the structure formation model is improved. The fiducial $\fnl$ value is marked by the dotted line.}
	\label{fig:resolution_fnl}
\end{figure}

The results in this paper demonstrate that future applications of field-level inference of primordial non-Gaussianity to galaxy redshift data will provide competitive constraints on $\fnl$. Here, we briefly discuss three necessary and/or possible extensions of the method. Firstly, future work will explore the modelling of scale-dependent bias parameters, $b_\phi$ and $b_{\phi,\delta}$, by relaxing the fixed assumptions based on the universal mass function \citep[][]{Barreira_2020b}, and sampling these parameters with and without priors. Investigations will also include alternative parametrisations, such as sampling $p$ or jointly sampling $\fnl \times b_\phi$, to assess their impact on $\fnl$ inference at the field level \citep{2024arXiv241018039S}. Secondly, the analysis can be extended to sample additional primordial cosmological parameters, such as $n_{\rm s}$ and $A_{\rm s}$, as well as other shapes of primordial non-Gaussianity, e.g. $f_{\rm NL}^{\rm equi}$ and $f_{\rm NL}^{\rm orth}$ \citep{Coulton_2023}. An extended analysis including these parameters could uncover parameter correlations with $\fnl$, and such degeneracies can affect sampling efficiency and parameter constraints. Third, increasing the resolution of the inference framework to include smaller scales, as suggested in Figure \ref{fig:resolution_fnl}, holds the potential for tighter constraints in $\fnl$, but requires improvements to the forward model, bias parameterisation, and noise parameter modelling $\sigma_{\rm G}$, to address nonlinearities and the breakdown of the Gaussian assumption. These advances aim to improve the accuracy of field-level inference on all scales and parameters \citep{2023ApJ...952..145J,2023arXiv231209271D,2024arXiv240807699J,2025MNRAS.542.1403D}, and could yield even tighter constraints on $\fnl$.

\section*{Acknowledgements}

The authors thank Nhat-Minh Nguyen, Fabian Schmidt, and Emiliano Sefusatti for insightful feedback on the manuscript.

We acknowledge the use of the following packages: \texttt{NumPy} \citep[]{harris2020array}, \texttt{Matplotlib} \citep[]{Hunter_2007}, \texttt{GetDist} \citep[]{lewis2019getdist}, and \texttt{HEALPix} \citep[]{gorski_healpix_2005}.\\

This work is done within the Aquila Consortium (\url{https://www.aquila-consortium.org/}).\\

This research utilised the HPC facility supported by the Technical Division of the Department of Physics, Stockholm University.\\

AA acknowledges with gratitude the financial support provided by the Wenner-Gren Foundations under Grant No. WGF2025-0043, and the grant 'Cosmology and Fundamental Physics with CMB and LSS' funded by Progetti di Astrofisica Fondamentale INAF 2023 and by the INFN InDark initiative, and the support from the Cosmology and Astroparticle Student and Postdoc Exchange Network (CASPEN), which facilitated collaborative visits.

JJ and GL acknowledge support from the Simons Foundation through the Simons Collaboration on "Learning the Universe". GL acknowledges support from the CNRS-IEA "Manticore" project.\\

The computations and data handling were enabled by resources provided by the National Academic Infrastructure for Supercomputing in Sweden (NAISS) and the Swedish National Infrastructure for Computing (SNIC) at Tetralith partially funded by the Swedish Research Council through grant agreements no. 2022-06725 and no. 2018-05973.\\
BDW acknowledges support from the \^Ile de France region in form of the DIM ORIGINES 2023 INFINITY NEXT grant. 
GJ acknowledges support from the ANR LOCALIZATION project, grant ANR-21-CE31-0019 / 490702358 of the French Agence Nationale de la Recherche. MM and ML acknowledge support by the MUR Progetti di Ricerca di Rilevante Interesse Nazionale (PRIN) Bando 2022 - grant 20228RMX4A, funded by the European Union - Next generation EU, Mission 4, Component 1, CUP C53D23000940006.
The Flatiron Institute is supported by the Simons Foundation.

\section*{Data Availability}

The data underlying this article will be shared on the basis of a reasonable request to the corresponding author.

\newpage

\bibliographystyle{mnras}
\bibliography{mnras}

\appendix

\section{Choice of the parameter $\lowercase{p}$}
\label{app:p_sdb}

In earlier tests, we performed the inference with $p=1.0$. In this configuration, we observed a mismatch in the inferred linear bias values between simulations with different $f_{\mathrm{NL}}$ values, despite identical initial conditions. In particular, the three simulations with $f_{\mathrm{NL}}=-100,\,0,\,100$ yielded systematically different linear bias parameters.

Following the discussion in \cite{2025arXiv250811798S}, we reduced the effective amplitude of the scale-dependent bias contribution. While that work achieves this by modifying $\delta_c$, we instead varied the parameter $p$, increasing it from $1.0$ to $1.3$.

When repeating the fixed-initial-condition tests with $p=1.3$, the offset between the inferred linear bias values was distrinctly alleviated. This indicates that the previously observed shift in the linear bias was driven by a mismatch in the modelled amplitude of the scale-dependent bias effect. For this reason, all inferences presented in the remainder of this work were carried out with $p$ fixed to $1.3$.

\section{Cross-correlation power spectrum}

The cross-correlation power spectrum $r_{i,j}(k)$ quantifies the correlation between two fields, $i$ and $j$, as a function of wavenumber $k$. It is defined as the ratio of the cross-power spectrum $P_{i,j}(k)$ to the geometric mean of the auto-power spectra $P_{i,i}(k)$ and $P_{j,j}(k)$, such that:

\begin{equation}
    r_{i,j}(k) = \frac{P_{i,j}(k)}{\sqrt{P_{i,i}(k) P_{j,j}(k)}}.
\label{eq:cross_correlation}
\end{equation}
Here, $P_{i,j}(k)$ represents the cross-power spectrum between fields $i$ and $j$, while $P_{i,i}(k)$ and $P_{j,j}(k)$ denote the auto-power spectra of fields $i$ and $j$, respectively.

The cross-correlation coefficient $r_{i,j}(k)$ ranges from $-1$ to $1$, where $r_{i,j}(k) = 1$ indicates perfect correlation at a given wavenumber $k$, $r_{i,j}(k) = -1$ indicates perfect anticorrelation and $r_{i,j}(k) = 0$ signifies no correlation.

\section{Summary of $\fnl$ results}

\begin{table}
\centering
\begin{tabular}{|c|c|c|c|c|c|c|}
\hline
Index & $\langle f_{\rm NL, 0} \rangle$ & $\sigma_{f_{\rm NL, 0}}$ & $\langle f_{\rm NL, +} \rangle$ & $\sigma_{f_{\rm NL, +}}$ & $\langle f_{\rm NL, -} \rangle$ & $\sigma_{f_{\rm NL, -}}$ \\
\hline
1 & -11.00 & 21.58 & 83.85 & 23.80 & -98.70 & 24.88 \\
2 & -23.74 & 22.65 & 58.53 & 24.52 & -96.56 & 25.85 \\
3 & 35.85 & 19.40 & 116.21 & 23.47 & -84.12 & 19.38 \\
4 & -11.14 & 20.79 & 76.50 & 23.28 & -102.82 & 25.26 \\
5 & 22.86 & 21.15 & 99.11 & 24.38 & -84.75 & 21.76 \\
6 & -23.96 & 21.98 & 68.87 & 24.19 & -111.22 & 23.80 \\
7 & 7.72 & 17.47 & 97.62 & 20.60 & -71.81 & 19.02 \\
8 & 17.27 & 22.09 & 116.29 & 25.05 & -91.40 & 23.44 \\
9 & -21.52 & 21.12 & 64.39 & 22.51 & -107.46 & 23.30 \\
10 & -28.46 & 19.77 & 67.50 & 20.79 & -128.18 & 24.67 \\
\hline
\end{tabular}
\caption{Summary of mean and standard deviation for $f_{\rm NL}$ in each inference, as shown in Figure \ref{fig:highlight}. The $_{0}$, $_{+}$, and $_{-}$, subscripts refer to the $\fnlgt=0,100,-100$ case, respectively.}
\label{tab:fnl_summary}
\end{table}

To summarise the combined statistics for the sets of $\fnl$ measured with uncertainties $\sigmafnl$, the weighted mean is given by $f_{\rm NL}^{\text{weighted}} = \frac{\sum_{i} w_i f_{\rm NL}^i}{\sum_{i} w_i}$, where $w_i = \frac{1}{\sigmafnl^2}$. The total standard deviation includes contributions from individual uncertainties and scatter relative to the ground truth $\fnlgt$: $\sigma^{\text{total}}(\fnl) = \sqrt{\frac{1}{\sum_{i} w_i} + \frac{\sum_{i} w_i (f_{\rm NL}^i - f_{\rm NL}^{\text{true}})^2}{\sum_{i} w_i}}$. In Table \ref{tab:fnl_summary}, we list all the inferred values and uncertainties of $\fnl$ plotted in Figure \ref{fig:highlight}.

In Table \ref{tab:fnl_summary}, we list all the inferred values and uncertainties of $\fnl$ plotted in Figure \ref{fig:highlight}.

\begin{figure}
	\centering
    \includegraphics[width=1.0\columnwidth]{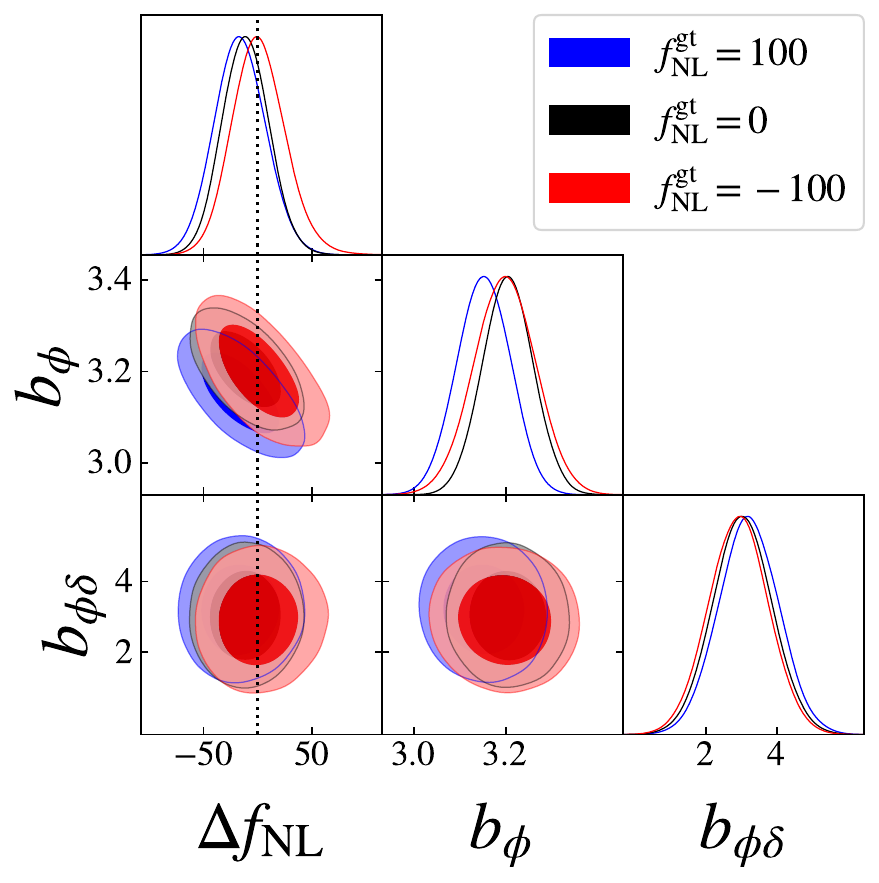}
	\caption{Corner plot showing the posterior distributions of $\fnl$, $b_{\phi}$, and $\bpd$. The values of $b_{\phi}$ and $\bpd$ are derived from the linear bias parameter $ b_1 $ and the second-order bias parameters $ b_1 $, $ b_2 $, respectively. The plot includes results from the first three analyses with ground truth values of $\fnlgt= 100$, $\fnlgt= 0$, and $\fnlgt= -100$. The corner plot is based on the same run as Figure \ref{fig:mr_corner}.
    }
    \label{fig:corner_bp_bpd}
\end{figure} 

As a visualisation of the constraints on $\fnl$, the effective $b_{\phi}$, and the effective $\bpd$, we show a corner plot of these three parameters in Figure \ref{fig:corner_bp_bpd}. The bias parameters $b_{\phi}$ and $\bpd$ are derived from the linear bias parameter $ b_1 $ and the second-order bias parameter $ b_2 $, respectively, as defined in Eqs \ref{eq:bp} and \ref{eq:bpd}. The figure illustrates the constraints on $b_{\phi}$ and $\bpd$ that \borg{} is able to measure from the data, under the assumed data model, which are directly relevant for the constraints on $\fnl$.

\begin{figure*}
	\centering
    \includegraphics[width=1.80\columnwidth]{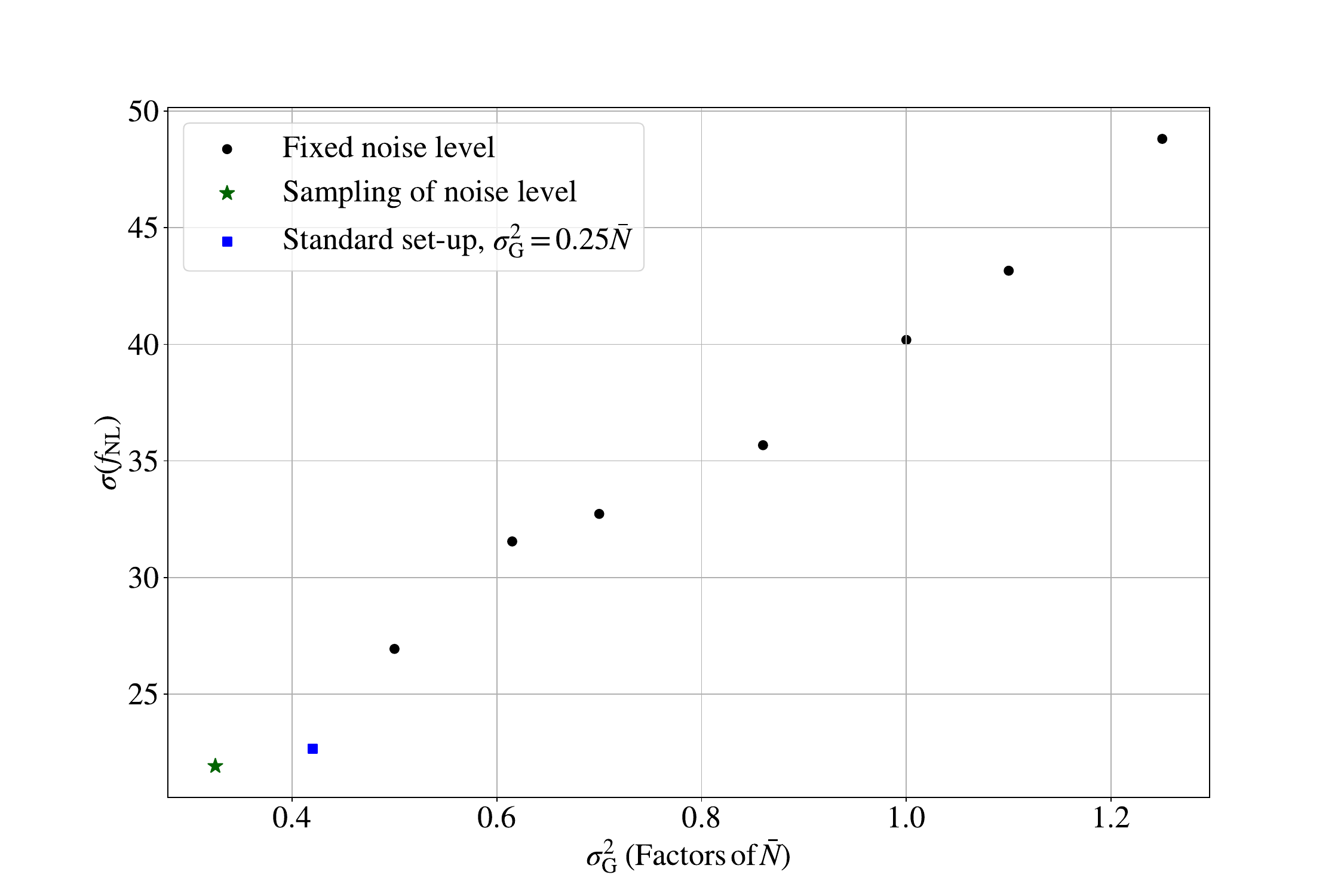}
	\caption{Relationship between $\sigma^2_{\rm G}$, expressed as a fraction of $\bar{N}$, and the uncertainty in $f_{\rm NL}$ ($\sigma(f_{\rm NL})$). The figure compares scenarios with fixed noise levels (black points) to a setup where $\sigma_{\rm G}$ is sampled (dark green star) and the standard configuration of this paper, where $\sigma_{\rm G}^2 = 1.2 = 0.42\bar{N}$ (blue square).}
	\label{fig:sigmag_sigmafnl}
\end{figure*} 

\begin{figure*}
	\centering
    \includegraphics[width=1.75\columnwidth, trim={0.0cm 0cm 0.0cm 0cm},clip]{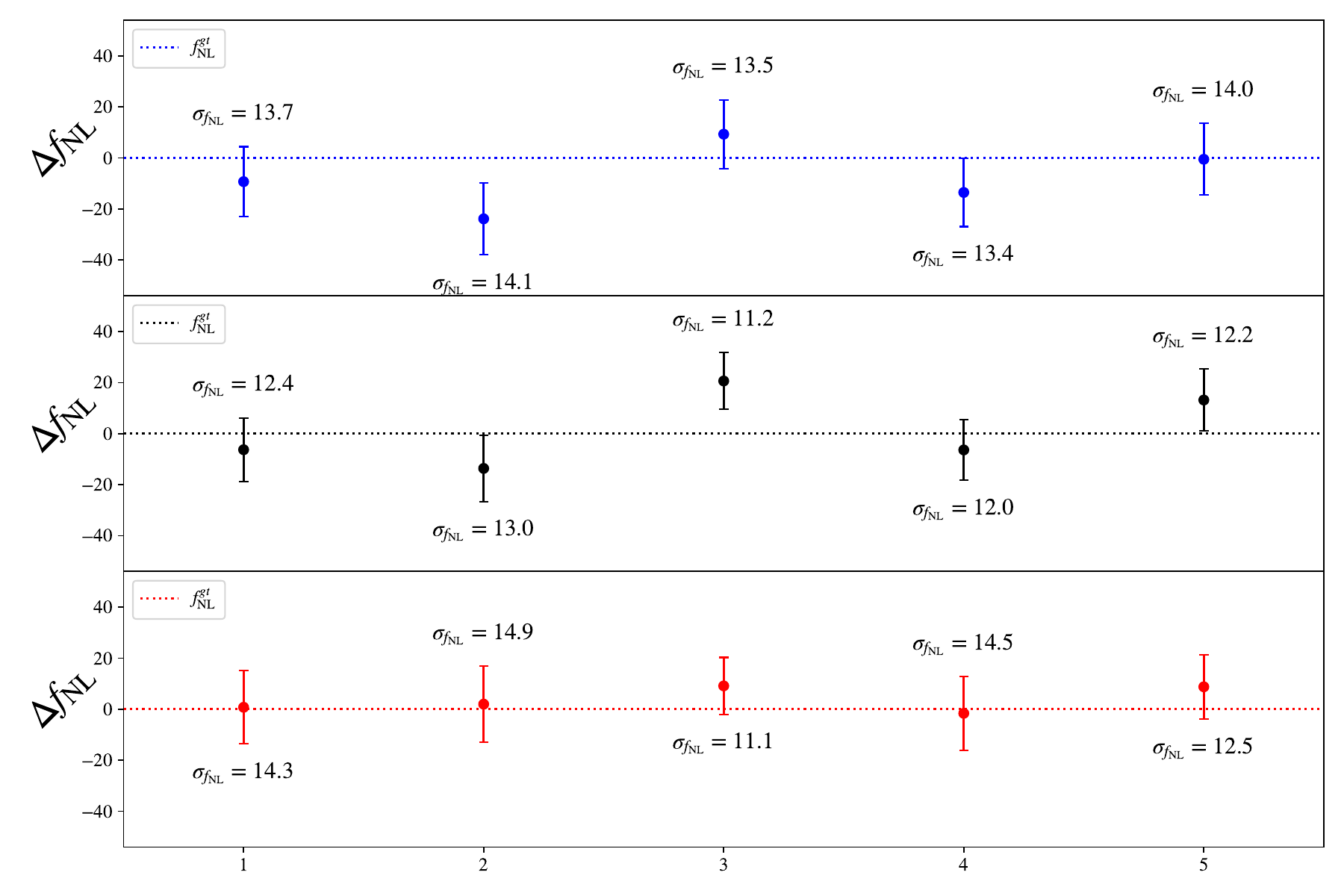}
    \caption{Comparison of inferred $f_{\rm NL}$ values over the 15 fixed-initial-condition analyses, divided into positive $\fnlgt$ (top row), null $\fnlgt$ (middle row), and negative $\fnlgt$ (bottom row). The error bars represent the uncertainty $\sigmafnl$, and the horizontal dashed lines represent the ground-truth $\fnlgt$ value. 
    }
	\label{fig:highlight_fic}
\end{figure*}

\section{Testing the impact of the likelihood noise level on inferred uncertainty on local primordial non-Gaussianity}
\label{sigmagsigmafnl}

We investigate the impact of the parameter $\sigma_{\rm G}$ on two fronts: Firstly, with a run where $\sigma_{\rm G}$ is sampled freely together with the initial conditions, bias parameters, and $\fnl$. Secondly, by fixing $\sigma_{\rm G}$ to larger values relative to the chosen value for the main runs of the paper (i.e. $\sigma^2_{\rm G}=1.2$). To maintain consistency, start from the same Markov state in a MCMC chain, for the same data set considered (the first one, with $\fnlgt=0$). The motivation for this is that we only want to test the impact on $\sigmafnl$ -- we note that this approach does not constitute a full investigation into all of the effects that changing $\sigma_{\rm G}$ has on the MCMC run.

The results are presented in Figure \ref{fig:sigmag_sigmafnl}. Our findings indicate a correlation between $\sigma_{\rm G}$ and $\sigmafnl$. Moreover, the figure shows that the resulting inferred $\sigma_{\rm G}$ is lower than the value we fix $\sigma_{\rm G}$ at. Importantly, at resolution $N=32$, with a linear bias model, we observe that $\sigma$ collapse occurs \citep{2021JCAP...03..058N}. However, using the bias model adopted in this paper (Eq. \ref{eq:bias_model}) appears to stabilise it. Overall, the results suggest that $\sigma_{\rm G}$ plays a role in the inferred uncertainty in $\fnl$, and motivate the need for further exploration in the choice of priors on $\sigma_{\rm G}$, particularly in the context of more realistic galaxy datasets where complex noise properties and observational effects come into play. Another effect of also sampling $\sigma_{\rm G}$ is the increase in the correlation length for the MCMC analysis \citep{2024arXiv240303220N}. A more complete study of this effect will be conducted in a future publication.

\section{Convergence testing}
\label{FULL_runs}

To assess convergence robustness in the first three simulations, we conducted a set of supplementary MCMC runs for each simulation. Each chain was initialised with different parameter values to ensure independence. After discarding the warm-up phase, during which the MCMC chains are allowed to move through the posterior space, we monitor convergence by computing the Gelman–Rubin (G-R) statistic \citep[][]{gr_test_1992}. Convergence is achieved if the G-R statistic approaches a value of $R\approx0.01$, indicating that the interchain variance aligns closely with the intrachain variance.

\begin{figure}
	\centering
    \includegraphics[width=1.0\columnwidth]{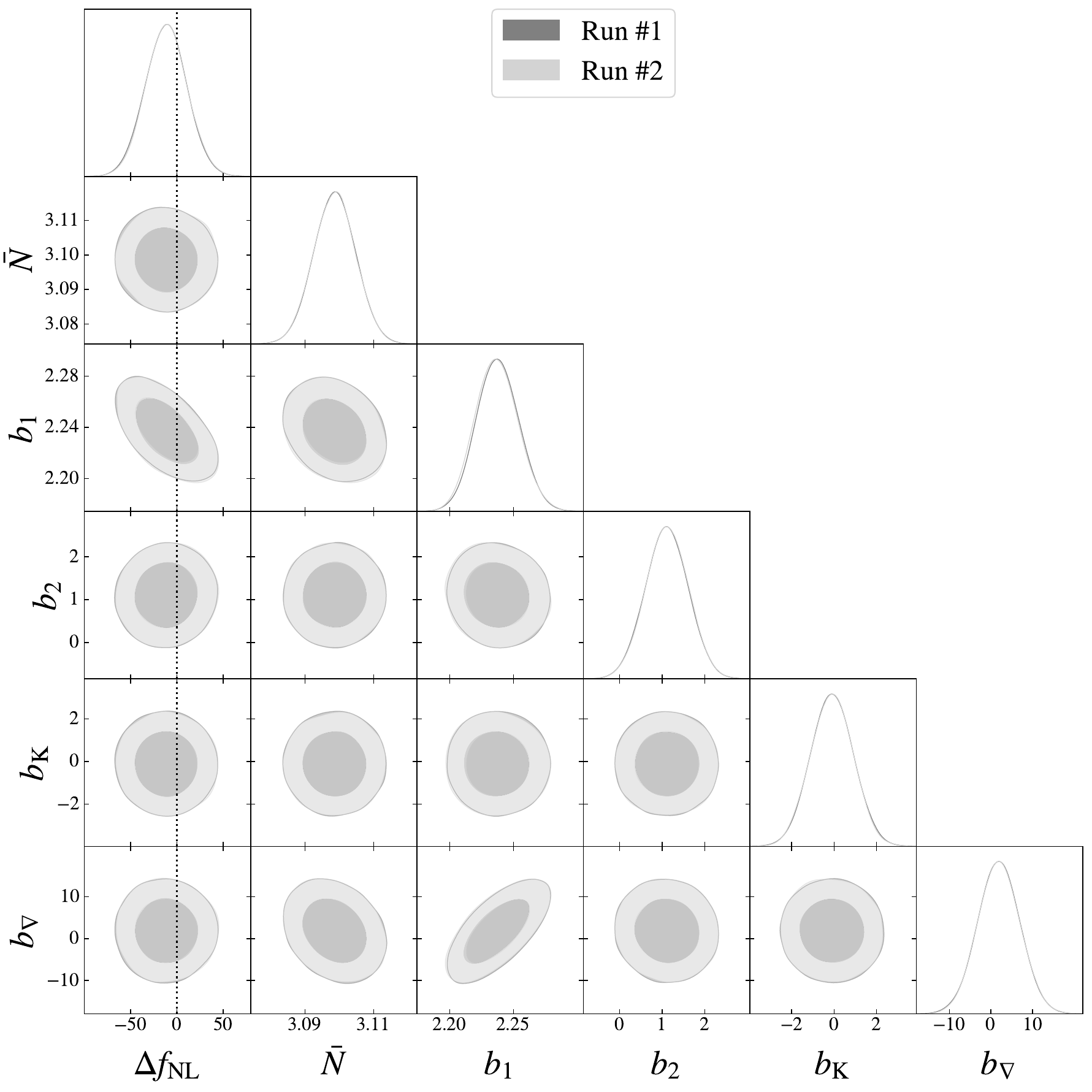}
	\caption{Marginalised posteriors of bias parameters and $\fnl$, for the two chains analysing the first simulation with $\fnlgt=0$. Initial conditions are sampled. The results indicate that the two chains are converging around the same values of the bias parameters and $\fnl$, displaying consistency in the inference.}
	\label{fig:corner_z1z2}
\end{figure}

\begin{figure}
	\centering
    \includegraphics[width=1.0\columnwidth]{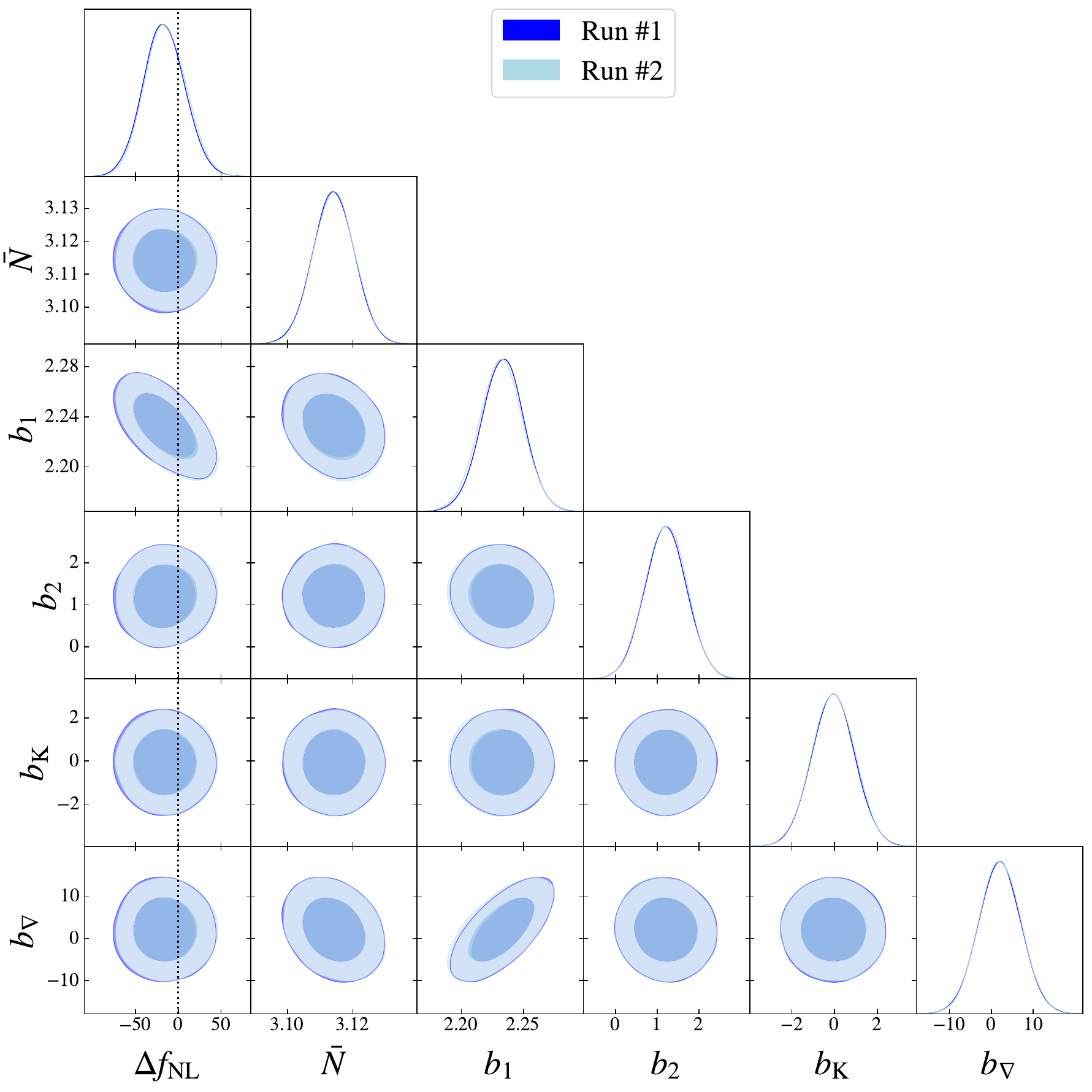}
	\caption{Same as Figure \ref{fig:corner_z1z2}, but for the first simulation with $\fnlgt=100$.}
	\label{fig:corner_p1p2}
\end{figure}

\begin{figure}
	\centering
    \includegraphics[width=1.0\columnwidth]{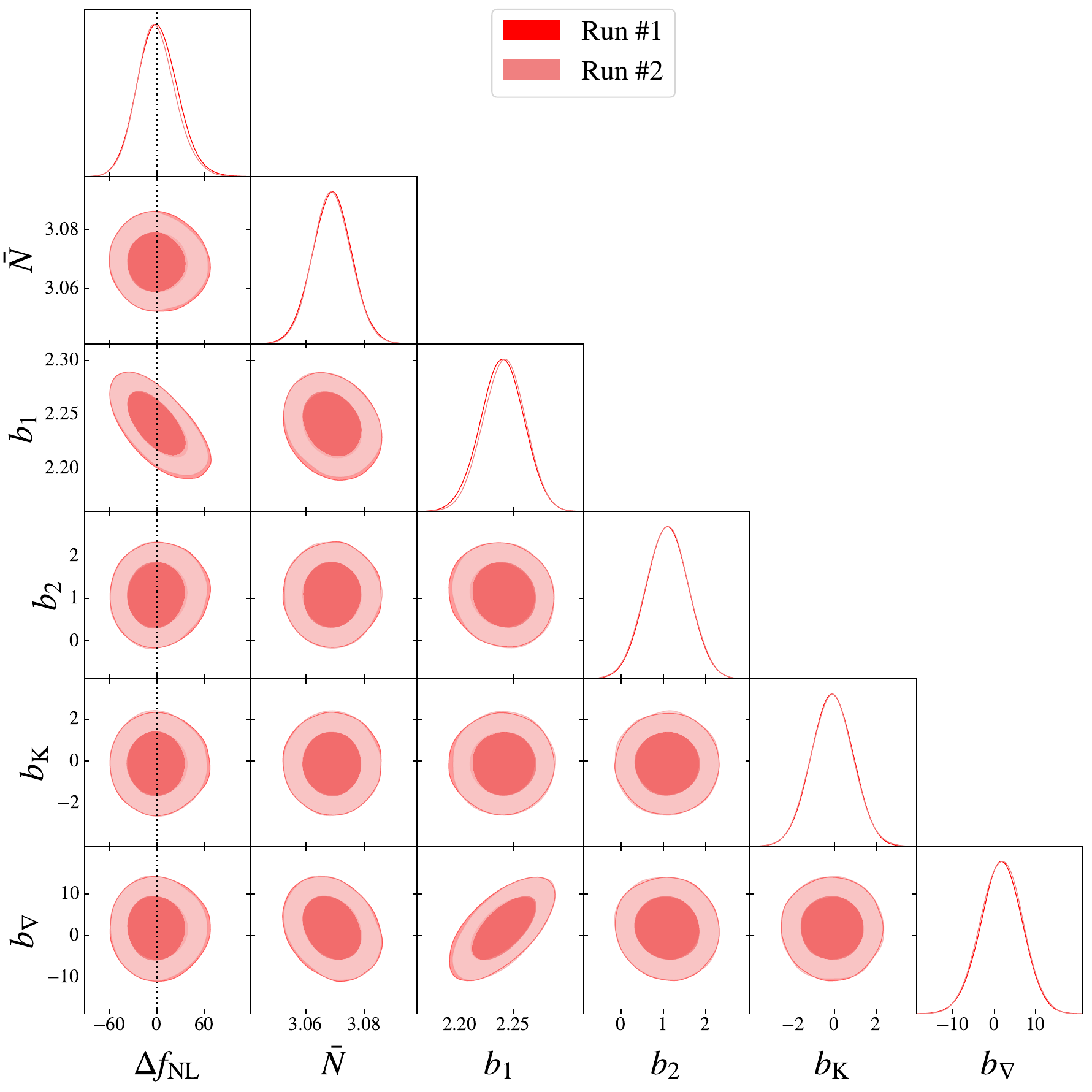}
	\caption{Same as Figure \ref{fig:corner_z1z2}, but for the first simulation with $\fnlgt=-100$.}
	\label{fig:corner_n1n2}
\end{figure}

As shown in Figures \ref{fig:corner_z1z2}, \ref{fig:corner_p1p2}, and \ref{fig:corner_n1n2}, the resulting posterior distributions of $\fnl$ and the bias parameters are visually consistent, supporting the convergence of each MCMC chain around the target distribution.

\section{\MakeLowercase{t}COLA summary}
\label{tcola_sect}

The tCOLA method provides a hybrid approach for simulating large-scale structure formation, combining analytic LPT with numerical particle mesh methods. The core idea is to divide the Lagrangian displacement field, $\Psi(q, a)$, into two components: $\Psi_{\text{LPT}}(q, a)$, the analytically calculated LPT displacement, and $\Psi_{\text{res}}(q, a)$, the residual displacement obtained from numerical particle mesh methods. The equations of motion governing the evolution of $\Psi_{\text{res}}(q, a)$ are
\begin{align}
\frac{\partial^2}{\partial a^2} \Psi_{\text{res}}(q, a) = -\nabla_x \Phi(x, a) - \frac{\partial^2}{\partial a^2} \Psi_{\text{LPT}}(q, a),
\end{align}
where $\Phi(x, a)$ is the gravitational potential that satisfies the Poisson equation. By evolving large scales using LPT and small scales using a limited number of time steps, tCOLA provides a computationally efficient solution that preserves the accuracy of LPT on large scales while approximating nonlinear small-scale dynamics. For more details, we refer the reader to \citet[][]{tassev_solving_2013}.

\label{lastpage}

\end{document}